\documentclass[submission,copyright,creativecommons]{eptcs}
\providecommand{\event}{WACA 2025} 

\usepackage{iftex}

\ifpdf
  \usepackage{underscore}         
  \usepackage[T1]{fontenc}        
\else
  \usepackage{breakurl}           
\fi

\usepackage{tikz}
\usepackage[normalem]{ulem}
\usepackage{amsmath}
\usepackage{booktabs}
\usepackage[]{algorithm2e}
\usepackage{amssymb}
\usepackage{caption}
\usepackage{subcaption}
\usepackage{arydshln}
\usepackage[english]{babel}
\PassOptionsToPackage{strings}{underscore}
\usepackage[strings]{underscore}
\usepackage{listings}
\usepackage{xcolor}

\providecommand{\thisvolume}[1]{this volume of EPTCS, Open Publishing Association}

\lstdefinelanguage{YAML}{
  morekeywords={true,false,null},
  sensitive=false,
  morecomment=[l]{\#},
  morestring=[b]',
  morestring=[b]"
}

\usepackage{appendix}

\newcommand*\circled[1]{\tikz[baseline=(char.base)]{
            \node[shape=circle,fill,inner sep=0.4pt] (char) {\textcolor{white}{#1}};}}

\title{Optimal Configuration of API Resources in Cloud Native Computing}

\author{\qquad Eddy Truyen  \qquad\qquad Wouter Joosen
\institute{DistriNet, KU Leuven, 3001 Leuven, Belgium}
\email{\qquad Eddy.Truyen@kuleuven.be \quad\qquad Wouter.Joosen@kuleuven.be}}

\def\titlerunning{Optimal Configuration of API Resources in Cloud Native Computing}
\def\authorrunning{E. Truyen \& W. Joosen}

\begin{document}

\maketitle

\begin{abstract}
This paper presents how an existing framework for offline performance optimization can be applied to microservice applications during the Release phase of the DevOps life cycle. Optimization of resource allocation configuration parameters for CPU and memory during the Release phase remains a largely unexplored problem as most research has focused on intelligent scheduling and autoscaling of microservices during the Ops stage of the DevOps cycle. Yet horizontal auto-scaling of containers, based on CPU usage for instance, may still leave these containers with an inappropriately allocated amount of memory, if no upfront fine-tuning of both resources is applied before the Deployment phase.
We evaluate the performance optimization framework using the TeaStore microservice application and statistically compare different optimization algorithms, supporting informed decisions about their trade-offs between sampling cost and distance to the optimal resource configuration. This shows that upfront factor screening, for reducing the search space, is helpful when the goal is to find the optimal resource configuration with an affordable sampling budget. When the goal is to statistically compare different algorithms, screening must also be applied to make data collection of all data points in the search space feasible.  If the goal is to find a near-optimal configuration, however, it is better to run bayesian optimization without screening. 
\end{abstract}


\section{Introduction}
\label{introduction}
\paragraph{Context}
Modern software increasingly adopts cloud-native deployment models, shifting from monolithic architectures towards distributed applications composed of containerized microservices. Such applications are typically deployed on container orchestration (CO) platforms, with Kubernetes as the most widely used example~\cite{kratzke2017understanding}. These platforms support DevOps practices and continuous integration/deployment (CI/CD), enabling rapid iteration and release~\cite{balalaie2016microservices}. 

A defining feature of Kubernetes (K8s) is that it exposes a wide variety of configuration parameters through its API. These \textit{API resources} include not only low-level infrastructure settings such as CPU and memory allocation parameters for containers, but also application-level parameters such as environment variables, replica counts, or string-based configuration options. Correctly configuring these resources is critical, as they directly influence application performance, cost efficiency, and compliance with service-level objectives (SLOs)~\cite{beyer2018site,schurman2009user}. 

We build on our existing framework \textit{k8-resource-optimizer}~\cite{kaminskiwoc2019,k8-resource-optimizer/impl-public} that applies black-box optimization algorithms to map application-level SLOs for a given K8s-based application to cost-efficient API resource configuration settings. Unlike manual approaches, black-box optimization treats the application as a system under test and requires no manually-specified performance model~\cite{alipourfard2017cherrypick,hsu2018arrow}. The framework can be instantiated with an off-the-shelf optimization algorithm, a benchmark scenario for a specific application, as well as an utility function for scoring how well an application configuration meets the optimization goal. Once these plugin components are instantiated, the user needs to provide minimal input: the manifest name of the application (e.g. a helm chart), the workload intensity, the SLOs to be met, and the set of tunable API resource parameters. The framework can be used to fine-tune \emph{any} resource parameter that is configurable via the K8s API, thus supporting  performance optimization of a wide range of numerical and string-based application configuration parameters. 

Yet the sheer number of configuration parameters exposed by the K8s API poses a significant challenge. It is not clear whether existing optimization algorithms can reliably navigate such an expansive space to find near-optimal configurations, especially under tight search and sampling budgets. Black-box optimization, in particular, faces two key challenges when applied to performance optimization of K8s-based applications. First, the cost of evaluating a single application configuration is high, as it requires deploying the application with specific settings and running statistically significant benchmarking tests. This severely limits the number of samples that can be afforded. Second, the size of the search space grows rapidly with the number of tunable API resources, many of which may have little or no influence on the optimization goal. 

This is where factor screening comes in. Originating in the design of experiments literature, factor screening~\cite{morris1991factorial} aims to identify which parameters of a function have the largest effect on function outcomes and which can be fixed at default values without significant impact. Applied to K8s, this means reducing the dimensionality of the search space by focusing optimization efforts only on those API resources that matter most. The benefits are twofold: (i) optimization algorithms can locate optimal or near-optimal configurations more cost-efficiently, i.e. using a lower number of samples, and (ii) when the reduced search space becomes small enough, exhaustively evaluating all samples becomes feasible, which in turn enables offline comparison between optimization algorithms without repeatedly running costly tests. Thus, factor screening acts as a bridge between the wide variety of API resource parameters exposed by K8s and the practical constraints of performance optimization.  

\paragraph{Goal}
To demonstrate the above ideas, this paper presents a case study in the context of container resource allocation only. We focus specifically on CPU and memory allocation parameters of microservices, which are central to application performance and cost in K8s environments. For even these two parameters we show that factor screening can drastically reduce the search space for a single microservice-based application and thus improve the cost-efficiency of black-box optimization algorithms. While our study centers on CPU and memory, the methodology is general and applicable to other categories of API resources exposed by K8s. 

Accurately translating high-level SLOs into efficient container resource configurations has been a hot topic of research in the past decade. Most CO platforms require manual translation, often leading to suboptimal outcomes like over-provisioning and under-utilization of resources, which increases costs~\cite{rodriguez2018container}. Case studies show significant over-provisioning in resource, such as T-Mobile’s containers having only 5.15\% memory utilization~\cite{lahmann2018container}, and most jobs in production clusters being over-provisioned~\cite{jyothi2016morpheus}. This problem is exacerbated by the complexity of modern microservice stacks, where tuning can take weeks \cite{li2018metis}. Unsurprisingly, a large body of research has focused on using machine learning and optimization techniques to improve runtime efficiency through intelligent scheduling and autoscaling~\cite{survey_buyya_2022}.

\begin{figure}
  \centering
  \includegraphics[width=0.50\linewidth]{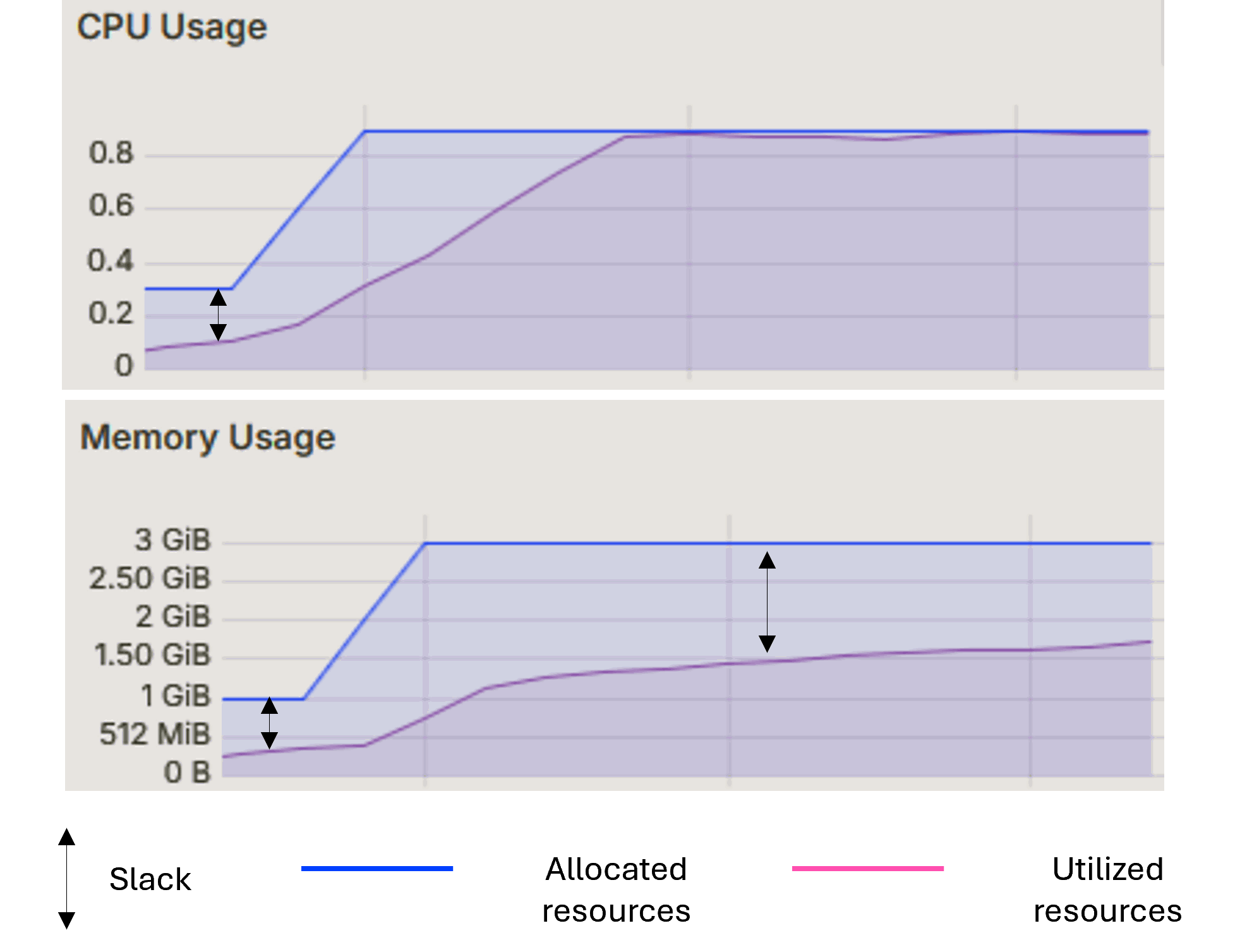}
  \caption{Example TeaStore trace: autoscaling triggered by CPU thresholds can leave memory over-provisioned. Therefore allocation of CPU and memory resources must be balanced offline during the Release phase.}
  \label{fig:motivation}
\end{figure}

However, most of this work addresses the operational stage of the DevOps lifecycle, i.e., after deployment~\cite{moreschini2025ai}. 
Runtime autoscalers adjust container replicas or resource limits in response to CPU or memory usage thresholds, but they do not guarantee optimal initial resource configurations.  As illustrated in Figure~\ref{fig:motivation}, horizontal auto-scaling based on CPU thresholds can still result in containers being over-provisioned with respect to memory. This gap between resource allocation and actual utilisation, termed \textit{slack}~\cite{autopilot}, even remains when autoscaling considers both CPU and memory metrics, since scaling is triggered when either metric crosses a threshold. Vertical auto-scaling~\cite{autopilot}, which adjusts the resource allocation of running containers, risks container restarts whenever the underlying programming language runtime is not able to auto-adjust to new resource limits. In addition, autoscalers typically operate on individual microservices and fail to capture cross-service dependencies. This motivates the need for \textit{offline optimization} in the Release phase, where resource configurations can be systematically explored under controlled workloads.

\paragraph{Contribution statement}
In this paper, we make the following contributions: \begin{itemize} \item We apply k8-resource-optimizer to offline resource optimization of the Teastore microservice benchmark application deployed on K8s. \item We show that upfront factor screening reduces the search space by $10^7$ for Teastore so that optimization algorithms have a higher probability to find the optimal configuration or to find a near-optimal configuration with a limited sampling budget. \item We demonstrate that k8-resource-optimizer can statistically compare different optimization algorithms to understand their trade-off between total sampling cost and distance to optimal resource configuration. \item We show how the framework can be extended beyond CPU/memory tuning to optimize application-level parameters, including feature toggles and other adaptation-relevant settings exposed through the K8s API. This provides a foundation for systematically deriving performance-aware adaptation rules during the Release phase. \end{itemize}

\paragraph{Overview of the paper} The paper is organized as follows. Section~\ref{motivation} presents the background to understand the remainder of this paper. Subsequently, Section~\ref{framework} describes the architecture and implementation of k8-resource-optimizer. Then, Section~\ref{evaluation} evaluates the framework in the context of the TeaStore microservice application. Next, Section~\ref{discussion} discusses the usefulness of k8-resource-optimizer for optimization of adaptation-relevant configuration parameters of microservices through the K8s API. Thereafter, Section~\ref{related_work} describes the related work on off-line performance optimization in cloud-native computing. Finally, Section~\ref{conclusion} presents our conclusions and directions of future work. 

\section{Background and motivation}
\label{motivation}
This section presents the overall approach behind performance optimization and introduces screening and specific optimization algorithms that are evaluated in the paper.

\subsection{Performance optimization}
\label{backgr:black-box}
 In performance optimization, the goal is to find an optimum of an utility function $f : X \rightarrow \mathbb{R}$ by iteratively sampling values of $f(x)$ in order to find a global optimum of $x_{opt} \in X$.  If evaluation of $f(x)$ is expensive, the search for the global optimum should be done efficiently in terms of the number of evaluations. Several optimization algorithms have been suggested to guide this search such as bayesian optimization, simulated annealing, and genetic algorithms~\cite{tuningstorage}. These algorithms typically make  trade-offs between the  \textit{exploration} of the search space and \textit{exploitation} of insights in already sampled regions to guide the selection of the next sample.
 
In what follows, we formalize the performance optimization problem for a given containerized microservice application as:
For a workload $W$, find the optimal or a near-optimal resource configuration $C^*(\Vec{x})$  that satisfies the $SLOs$ of given $SLA$ $s$ and minimizes the operational cost $P(\Vec{x})$. We use $\Vec{x}$ to denote the selected parameter settings of a configuration $C$, where $\Vec{x}$ includes resources settings (CPU, RAM, disk I/O, network bandwidth) for each of microservices part of the applications. Let $SLI(\Vec{x})$, Service Level Indicator~\cite{beyer2018site}, be the actual measurements of the performance test for settings $\Vec{x}$. Since, the definition for an optimal configuration depends heavily on the application context, we allow for a user-defined \textit{utility function} with configuration settings $\Vec{x}$ and $SLI(\Vec{x})$ as input parameters. Our goal is two-fold: Firstly, to find the configuration $C^*$ with utility score $u^*$ which has the smallest possible distance to the optimal configuration $C_W$ for the given workload. The distance between $C_W$ and $C^*$ is expressed as the absolute value the difference in utility $|u^* - u_W|$. Secondly, the time to find an optimal configuration is limited. Therefore, the search should be performed within a limited amount of performance test samples. 

\subsection{Optimization algorithms}
\label{optimizers-explained}

\noindent\textbf{Bayesian optimization} (BO) ~\cite{shahriari2016taking} has become a popular technique to solve optimization problems. It sequentially models the utility function as a stochastic process with configuration settings as input parameters. First, the \textit{configuration-vs-utility} space is modeled by regressing a set of prior sample points (configurations already evaluated). This allows to create an estimate of the entire search space, BO calculates a confidence interval of the utility function. This model is then sequentially refined by the addition of samples via Bayesian posterior updating. The next sample point is selected using a predefined \textit{acquisition function}. This function aims to balance exploitation (i.e., regions with a high probability of containing an optimum) and exploration (i.e., regions with high uncertainty).  We selected the popular Expected Improvement (EI) function for this task, as it has shown promising results in similar contexts~\cite{alipourfard2017cherrypick, li2018metis} and it does not require tuning of its own parameters~\cite{snoek2012practical}. BO's sequential approach is suitable for scenarios where the evaluation of a single data point is a time-consuming task~\cite{li2018metis}, hence it is a suitable approach for our use-case.

\noindent\textbf{BestConfig}~\cite{zhu2017bestconfig} explores the search space by iterations of their proposed divide-and-diverge (DDS) sampling method and recursive bound-and-search (RBS) algorithm. DDS is a form of stratified sampling to achieve a wide coverage of the search space (exploration). Based on  the results of samples proposed by DDS, RBS either narrows the search space near high-achieving samples (exploitation) or backtracks to avoid local optimums.

\begin{figure}
\centering
\begin{minipage}{0.48\textwidth}
  \centering
  \includegraphics[width=\textwidth]{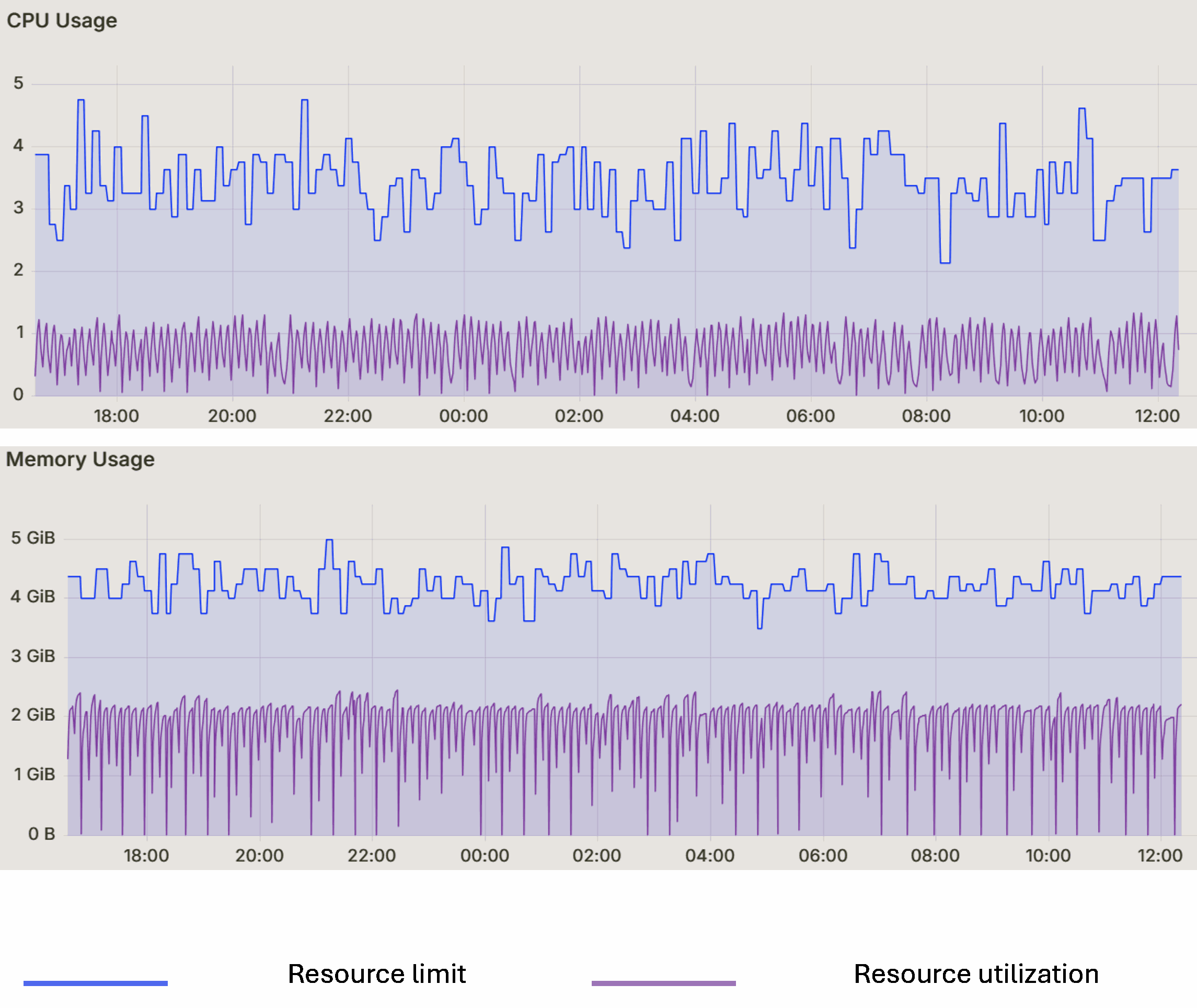}
  \caption{A single run of the BO algorithm taking 150 samples does not find a near optimal resource configuration for the TeaStore microservice application}
  \label{fig:bayesian_motivation}
\end{minipage}
\hfill
\begin{minipage}{0.48\textwidth}
    \centering
    \includegraphics[width=\textwidth]{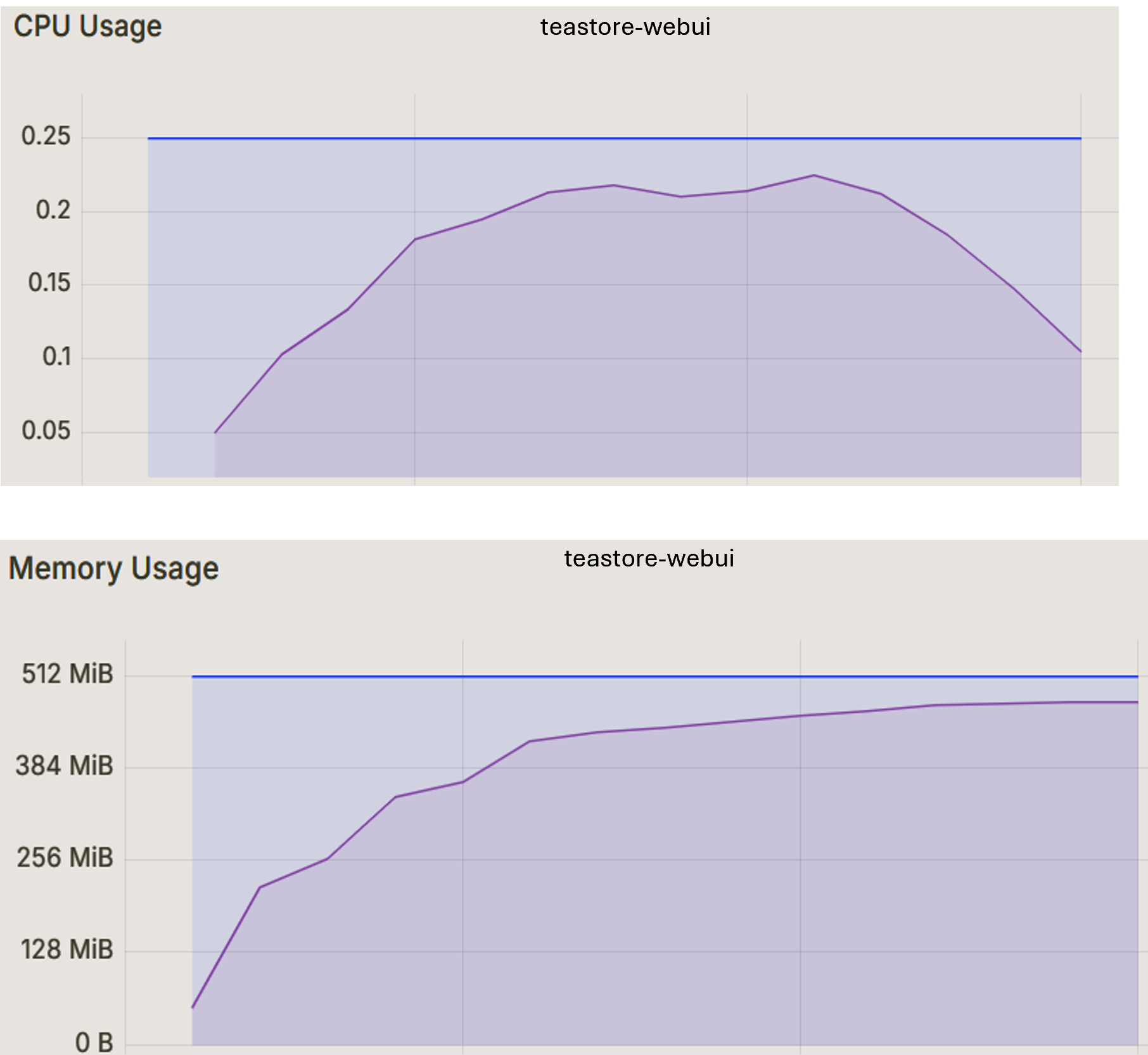}
    \caption{When setting TeaStore's resource configuration to the absolute lowest bound of the search space, performance bottlenecks do appear}
    \label{fig:bottleneck_motivation}
\end{minipage}
\end{figure}

Despite improved search algorithms, finding a near-optimal solution in a large search space may still require more samples than can be practically afforded. For example, Figure \ref{fig:bayesian_motivation} shows a resource monitoring screenshot of 150 configurations of the TeaStore application as selected by the BO. As shown, there is almost no improvement in the reduction of slack between allocated resources and utilized resources. Yet when we test the TeaStore application configuration that corresponds with all resource parameters set at the minimum of the search space, the WebUI microservice is a clear performance bottleneck (see Figure \ref{fig:bottleneck_motivation}). Finding all the bottlenecks in an application would help pinpoint those resource parameters that should be the locus of optimization.

\subsection{Factor screening}
\label{sec:meth:subsec:sens}
The size of the search space is determined by the search ranges for each parameter $[min, max]$. However, as developers might be clueless on resource requirements of their services, this search space may contain regions that are incapable of even remotely satisfying the SLO, and as such exploration of these regions would be pointless. Therefore it makes sense to reduce the full parameter space to a sub-area that has a high probability to contain the optimal resource configuration and that has a smooth performance surface. We assume that this allows off-the-shelf optimization algorithms to perform much better. With this end in view, it is possible to run k8-resource-optimizer with a factor screening/sensitivity analysis algorithm that relies on calculating Elementary Effects ($EE$) as implemented by Morris One-At-a-Time (MOAT)~\cite{morris1991factorial}. When applied to performance optimization, the MOAT algorithm determines which resources parameters have the most influence on the application performance for a certain benchmark scenario. Indirectly, it also indicates which application components may be performance bottlenecks.

The MOAT algorithm discovers for a given function which inputs have a substantial influence on the outputs. When applied to performance optimization,the function maps the resource parameters of the search space to attained Service Level Indicators (SLIs) such as latency and throughput, and the algorithm works as follows. The $k$-dimensional search space (for $k$ resource parameters), is partitioned in $p$ uniform levels for each parameter, resulting in a $p^k$ grid of possible settings. The method iteratively evaluates configurations in the search space, starting from a random point in the grid. 
During each iteration, one resource parameter $x_{i}$ is altered at a time in a discrete parameter space while fixing the other parameters. As such, a \textit{trajectory} is created in the search space of $k+1$ evaluations. Altering parameter $x_{i}$ in consecutive evaluations allows to calculate the elementary effect ($EE$) of that parameter, $EE_{i} = \frac{y(x_{1},...,x_{i}+\Delta,...,x_{k}) - y(x_{1},...,x_{i},...,x_{k})}{\Delta}$, where $y(x)$ is based on the Service Level Indicators (SLIs) obtained from the evaluations. We use $\Delta = \frac{p}{2(p-1)}$, leading to steps slightly larger than half the range of the normalized parameter range between $[0,1]$, as recommended by~\cite{saltelli2008global}. Repeating the process for $r$ \textit{trajectories} with r being between 5 to 15 (i.e., a total of $r*(k+1)$ evaluations), allows for the calculation of the mean $\mu_{i}$, modified mean $\mu_{i}^*$ (which are the absolute values of $EE_{i}$) and standard deviation $\sigma_{i}$ of each parameter $x_{i}$. The mean and modified mean express the overall influence of a configuration parameter on the SLIs, and standard deviation indicates dependence on other parameter settings~\cite{saltelli2008global}. 

Based upon these results of the screening algorithm, the parameter search ranges are adapted as follows: (i) the lower bound of a parameter is set to the minimum setting of that parameter tested capable of satisfying a more relaxed SLO (e.g., a tail-latency that is 40\% higher than the targeted SLO), (ii) the upper bound is set to the maximum setting of that parameter capable of satisfying a more strict SLO (e.g. a tail-latency that is 40\% lower than the targeted SLO), scaled by parameter's relative influence of $\mu_{i}^*$ as compared to the $\mu_{i}^*$ of all the other parameters. Formally, we define the lower and  upper bounds for parameter~$i$ as

\begin{equation}
    \mathrm{MinBound}_i \;=\; 
    \min_{x\in \mathcal{X}_i^{\text{relaxed}}} x,
\end{equation}
where $\mathcal{X}_i^{\text{relaxed}}=\{x\in \mathcal{X}_i:\ \text{SLO}_{\text{relaxed}}\ \text{is satisfied at setting }x\}$ 
is the subset of tested settings for parameter~$i$ that satisfy a relaxed SLO (e.g., tail latency $\leq 1.4\times$ the target SLO). 

We define the upper bound for parameter~$i$ as
\begin{equation}
    \mathrm{MaxBound}_i \;=\; \mathrm{MinBound}_i \;+\; 
    \Big(\max_{x\in \mathcal{X}_i^{\text{strict}}} x - \mathrm{MinBound}_i\Big)\,\rho_i,
\end{equation}
where $\mathcal{X}_i^{\text{strict}}=\{x\in \mathcal{X}_i:\ \text{SLO}_{\text{strict}}\ \text{is satisfied at setting }x\}$ 
is the subset of tested settings for parameter~$i$ that satisfy a stricter SLO (e.g., tail latency $\leq 0.6\times$ the target SLO). 

The scaling factor $\rho_i$ for parameter~$i$ is obtained by applying min--max scaling 
to its modified mean $\mu_i^*$ across all $k$ parameters:
\begin{equation}
    \rho_i \;=\; 
    \frac{\mu_i^* - \min_{1 \le j \le k} \mu_j^*}{\max_{1 \le j \le k} \mu_j^* - \min_{1 \le j \le k} \mu_j^*},
\end{equation}
so that $\rho_i \in [0,1]$, with the parameter having the lowest modified mean mapped to~0 
and the one with the highest mapped to~1.

In the degenerate case where all modified means are identical 
(i.e., $\max_{1 \le j \le k} \mu_j^* = \min_{1 \le j \le k} \mu_j^*$), 
we define
\begin{equation}
   \mathrm{MaxBound}_i \;=\; \min_{x \in \mathcal{X}_i^{\text{strict}}} x.
\end{equation}

\section{Framework}
\label{framework}
In this section, we present the command line interface, architecture and implementation of the k8-resource-optimizer framework.

\subsection{Command line interface}
\label{subsec:ui}
To optimize the deployment of a containerized microservice-based application on K8s, k8-resource-optimizer relies on two key inputs: (1) a helm chart that defines the base deployment including all K8s manifests; more specifically a \texttt{values.yaml} file inside the helm chart defines all the customizable K8s resource parameters and default values for these parameters, and (2) an optimizer configuration  file that specifies the service-level objectives (SLOs), and the parameter search space for tunable parameters of the Helm chart. In particular the optimizer configuration specifies one or more SLAs where each SLA presents a specific helm chart with a specific application. When more SLAs are specified, their corresponding applications are expected to be co-located on the same K8s cluster. Each SLA defines the following items: 
\begin{itemize}
    \item The name of the helm chart that should be used for deploying the application.
		\item One or multiple SLOs such as throughput or latency.
		\item The expected workload intensity in terms of the number of concurrently running tenants and the expected throughput per tenant.
    \item A set of tunable parameters for each microservice that must be defined upfront in the \texttt{values.yaml} file of the referenced helm chart. For each tunable parameter, a search range is defined (\texttt{min}, \texttt{max}, and \texttt{granularity}) and unit suffixes (e.g., \texttt{m} for millicores, \texttt{Mi} for mebibytes).
\end{itemize}

\begin{figure}
  \centering
  \includegraphics[width=0.75\linewidth]{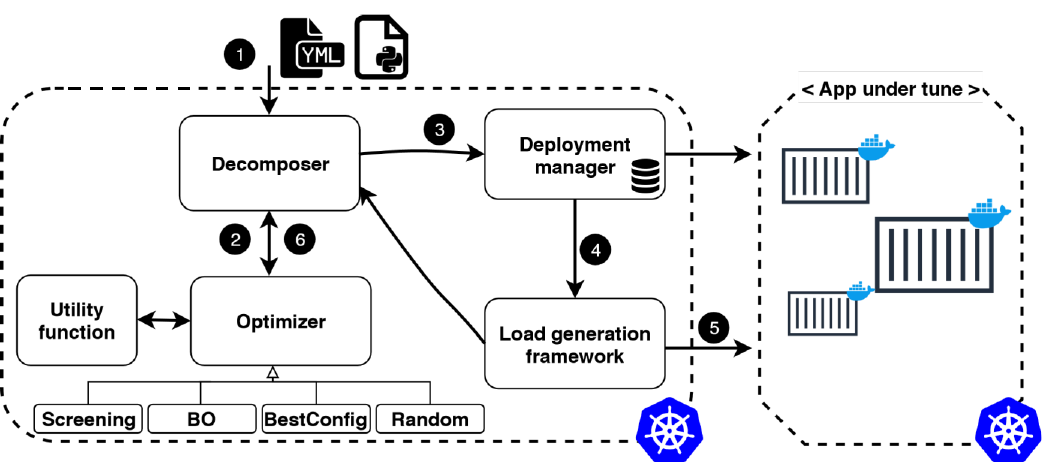}
  \caption{The architecture of k8-resource-optimizer.}
  \label{fig:architecture}
  
\end{figure}
		
Next an optimization strategy is included which consists of the following items:
    \begin{itemize}
        \item \texttt{nbOfIterations} and \texttt{nbOfSamplesPerIteration}, which control the search budget,
        \item \texttt{namespaceStrategy} (e.g., \texttt{NSPSLA} or \texttt{NSPT}) to determine if the application is multi-tenant or needs to be launched separately for each tenant.
        \item \texttt{optimizer} (e.g., \texttt{MOAT}, \texttt{bestconfig}) which selects the underlying optimization algorithm,
        \item \texttt{utilFunc}, a utility function used to evaluate the performance efficiency of each configuration.
    \end{itemize}

An example \texttt{values.yaml} file and optimizer configuration file for the TeaStore application are presented in Appendix \ref{sect:app-values} and \ref{sect:app-optimizer-configuration}, respectively.
K8-resource optimizer can also be used for optimization of any other resource parameter in the K8s API. For example, configuration parameters of the microservices can be optimized by making each parameter templatable in the Helm chart. This is a matter of including a variable for it in the \texttt{values.yaml} file. String based parameters should be converted in the helm chart to an Integer range with a granularity of 1. An example of the latter is shown in Appendix \ref{app:string-param}.  

Finally, a benchmark scenario is to be implemented when running k8-resource-optimizer by instantiating the \texttt{Experiment} interface. For microservice applications, we target an end-to-end user scenario where a user sequentially invokes multiple operations of the microservice application to simulate a full user journey. Such end-to-end user scenario is preferred as the basis of performance testing during optimization as this implicitly accounts for the dependencies between different microservices and optimizes the application as a whole instead of optimizing each microservice individually.

\subsection{Architecture}
The architecture of k8-resource-optimizer is illustrated in Figure~\ref{fig:architecture} and is composed  of loosely coupled components interacting via common interfaces. The data-flow between these components during optimization is given in Algorithm~\ref{alg:dataflow}. 

\begin{algorithm}[]
$\Rightarrow$ \circled{1} load optimizer configuration and performance test\;
 \For{iteration}{
  \circled{2}~ask Optimizer for next configuration $C(\Vec{x})$ to sample\;
    \circled{3}~deploy application with resource parameter settings of $\Vec{x}$\;
    \circled{4}~wait for application to be marked ready\;
    \circled{5}~obtain $\text{SLI}_1(\vec(x))$, $\text{SLI}_2\vec(x))$, ..., $\text{SLI}_n(\vec(x))$ from performance test\;
    \circled{6}~feed results to Optimizer\;
 }
 $\Leftarrow$ (near)-optimal configuration 
 \caption{Iteration loop of k8-resource-optimizer}\label{alg:dataflow}
\end{algorithm}

\paragraph{Optimizer}
The optimizer provides a reusable interface suitable for most optimization algorithms based upon exploration and exploitation. The predefined \textit{utility function} interface, described in Section~\ref{backgr:black-box}, allows the combination of any optimization algorithm and user-defined utility function.\\ 

\paragraph{Deployment Manager}
Responsible for the deployment of a particular helm chart and values.yaml file inside a K8s cluster. K8-resource-optimizer leverages K8s' readiness probes, to ensure that the application is fully deployed before running a performance test and retries a failed installation/removal of deployments.\\

\paragraph{Load generation}
In order to support a variety of load-generation frameworks (e.g., Locust, JMeter, Vegeta), k8-resource-optimizer provides an interface for experiments to be run as a performance test. Currently, the popular load-generation framework Locust~\cite{LocustHeyman} is integrated.

\subsection{Implementation}
\label{subsec:optimizer-implementation}
The core framework without plug-in components has been implemented using the Go programming language in 1.5K lines of code. To extend the k8-resource optimizer with a benchmark scenario and utility function for the TeaStore application, we need to implement an \texttt{Experiment} class and an \texttt{Utility\-Function} that together require 302 lines of code. We have implemented 11 different optimization algorithms including BO, BestConfig, MOAT, as well as random search and an exhaustive search algorithm in 2.9K lines of code. 
Finally, we also developed 1K lines of code for static analysis of the performance of different optimizations techniques to understand the optimality versus the search cost ratio of the these techniques.

\subsection{Setting workload intensity and resource parameter conversion}
 As stated in Section \ref{subsec:ui}, k8-resource-optimizer requires manual specification of the workload intensity as well as the parameters to be optimized and their conversion to K8s API resources.

The workload intensity is specified in terms of the number of concurrent tenants and a request or job arrival rate per tenant. How to set the workload intensity depends on the expected workload fluctuation pattern, i.e.,  bursty, monotonically increasing, seasonal and stable~\cite{nikravesh2015towards,lorido2014review}.  For stable and bursty workloads, the workload intensity should be set slightly larger than the constant workload volume of the stable workload and set equal to the highest expected peak for the bursty workload. For monotonic and seasonal workloads, one should set the highest expected workload intensity.

The conversion of parameters to K8s API resources relies heavily on the scripting facilities of helm charts. The \texttt{values.yaml} file of a helm chart typically specifies for each container one parameter per resource type and a parameter for the number of container replicas (cfr. Appendix \ref{sect:app-values}). The K8s \texttt{yaml} files in the helm chart then convert these parameters to K8s API resources as follows. 

Resource parameters in the \texttt{values.yaml} file must be converted to the K8s API \texttt{request} and \texttt{limit} entries. The \texttt{request} entry specifies the amount of node resources that should be available to the container, while the \texttt{limit} field sets the maximally allowed resource usage of the container. Both parameters are enforced by the operating system of the node~\cite{CloudNativeComputingFoundation}.  

In our experience, the resource parameters should be directly copied into the \texttt{limit} entry. The \texttt{request} entry's values are set equal to or a certain percentage lower than those of the \texttt{limit} entry~\cite{CloudNativeComputingFoundation,verma2015large}.  For stable workloads, this percentage will be smaller than for bursty workloads; if the application concerns a high-priority workload, however, requests must be equal to limits. 

Moreover, in stable or bursty workloads, the number of container replicas is fixed and typically set higher than 1 to ensure availability of the microservices. For monotonic and seasonal workloads, the number of microservice replicas is to be included as a tunable parameter in the optimizer configuration file of k8-resource-optimizer, and resource parameters are to be converted as in bursty workloads. The optimal number of replicas found should then be set as the maximum number of replicas in auto-scaling solutions.

\section{Evaluation}
\label{evaluation}
In this section, we illustrate the workflow of k8-resource-optimizer for TeaStore~\cite{von2018teastore}, a popular reference microservice application for performance evaluation of research prototypes. Thereafter we evaluate the usefulness of factor screening and compare the cost-effectiveness of the different optimization algorithms.

First, the experimental settings are explained in Section \ref{subsec:expenv}. Then factor screening is illustrated in Section~\ref{subsec:eval:sens}, providing insights into the obtained reduced search spaces for TeaStore v1.2.0. Next, Section~\ref{subsec:eval:exhaustive} shows the results of the exhaustive data collection process after factor screening, while Section~\ref{subsec:eval:comparative} evaluates to which extent the different optimization algorithms can find the optimal configuration as found by the exhaustive search, and Section~\ref{subsec:distance} compares the cost-effectiveness of different algorithms with respect to the trade-off between search cost and distance of a found near-optimal configuration to the optimal one. Subsequently, Section~\ref{subsec:screening} evaluates the usefulness of factor screening. Finally Section ~\ref{subsec:resource-utilization} evaluates the effective resource utilization of TeaStore as achieved by k8-resource-optimizer, hereby pointing to a few oversights in the overall workflow.

\subsection{Experimental environment}
\label{subsec:expenv}
This section provides a detailed description of the experimental environment and the configuration benchmark scenario used when running k8-resource-optimizer for TeaStore.

\paragraph{Infrastructure}
Our testbed consisted of a private OpenStack cloud. The physical machines of this cloud come with 2.60 GHz Intel Xeon E5-2660 processors and 128GB DDR3 memory. The K8s (v1.14) cluster consists of two nodes, master and worker, running on top of Ubuntu 16.04 VMs. The master has 4 CPU cores and 8GB of RAM and the worker node has 8 CPU cores and 16GB of RAM. These nodes are actually VMs that run on the same physical machine but CPU pinning is enabled to minimize performance interference. As such, physical CPU cores are exclusively reserved for a single VM and all virtual CPUs of a VM map to CPU cores that belong to the same motherboard socket.

\paragraph{Microservice application}
TeaStore~\cite{von2018teastore} is  a microservice application developed as a reference application for performance testing. It offers deployment specifications for multiple CO platforms where each microservice is deployed in its own container.  

\paragraph{Benchmark scenario and workload intensity}
For TeaStore, a number of requests are sequentially invoked as follows: login → get categories → view products in a category → add product to cart → view profile → logout. The full benchmark scenario is presented in Appendix \ref{app:scenario}. 
 
As workload intensity we let 10 concurrently running tenants execute the benchmark scenario. We assume a stable workload without bursts and therefore set the K8s resource requests equal to K8s resource limits. The tested SLO in the evaluation concerns the 99th percentile latency of separate requests, which is set at 1000 ms.

\paragraph{Utility function}
In our experiments we use the following utility function,
\[
    u(\Vec{x})= 
\begin{cases}
    1 + (SLI(\Vec{x}) - SLO)& \text{if } SLI(\Vec{x}) > SLO\\
      normP(\Vec{x})          & \text{otherwise}
\end{cases}
\]

where $normP(\Vec{x})$ equals to the normalized resource allocation cost of $\Vec{x}$. This utility function assigns a configuration that violates the SLO a score larger than 1 depending on the distance between the 99th percentile of the SLI and the SLO. It assigns configurations that satisfy the SLO a score between $[0,1)$, but discards their observed SLI in preference for resource allocation. The SLO-satisfying configuration with the lower resource allocation cost receives a lower score. The goal is to minimize this score.

\paragraph{Optimization algorithms}
For the implementation of BO we rely on a popular open-source Python library~\cite{bofmfn}. Here, the corresponding BO plugin maintains and communicates with a running process of the library. The library's specific implementation treats the search space as continuous. In our use-case, however, parameters have a certain granularity (e.g., 125 Mil), resulting in a discrete search space. Our optimization algorithm  therefore, translates the suggested samples by the BO to the nearest discrete settings. For BestConfig's optimization algorithm and the factor screening algorithm MOAT, we implemented the algorithms as described in their respective papers. 
Other baseline optimization algorithms are random search and incremental random search (RandomInc). It is an optimizer that shuffles the order of parameters randomly, and then systematically iterates over all possible configurations built incrementally based on that shuffled parameter order. In practice, it generates all configurations but only tests a subset, depending on sampling budget, meaning it is not truly exhaustive unless allowed to complete.

\subsection{Factor screening}
\label{subsec:eval:sens}
The first step of k8-resource-optimizer's methodology uses MOAT (cfr. Section~\ref{sec:meth:subsec:sens}) to determine reasonable parameter ranges and hereby reducing the search space before employing the optimization algorithms. The initial bounds for the parameter ranges for CPU and memory are set the same for all the TeaStore microservices. A minimum bound has been set by starting from the maximum bound and increasingly decreasing it with a granularity of 125 millicores for CPU resources and with a granularity of 128 Mi for memory resources, until the application does not start-up successfully. 
This results in a search space of $6^{14}$ possible configurations for TeaStore's 14 parameters, with each parameter having 6 possible settings. We run MOAT with 10 \textit{trajectories} resulting in 150 configurations to be evaluated for TeaStore v1.2.0. 

\begin{figure}
\centering
\begin{minipage}[t]{0.35\textwidth}
    \centering
    \includegraphics[width=\textwidth]{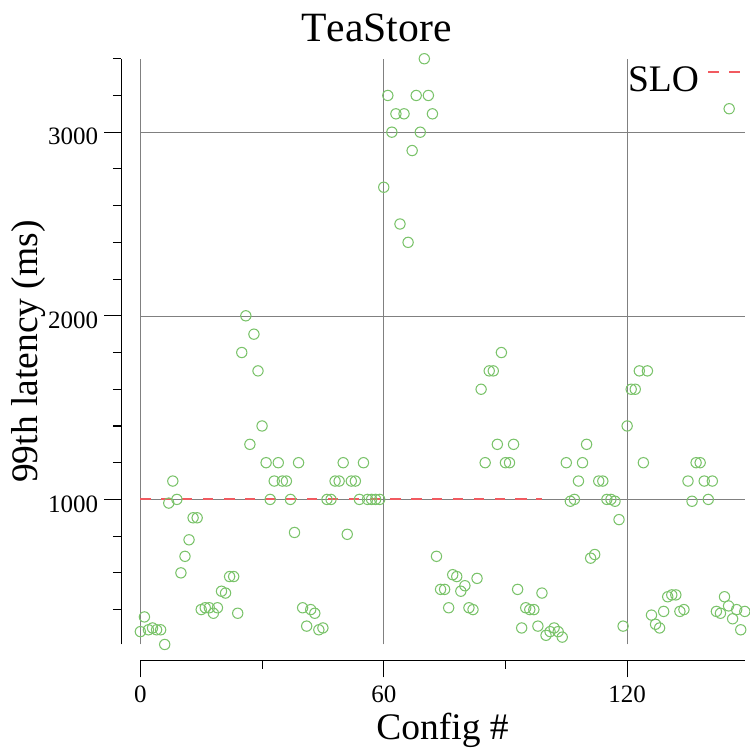}
    \caption{99th percentile latency for all configurations evaluated during factor screening. The data points are sorted by the order of evaluation of configurations by the screening algorithm. Clusters indicate low-impact parameters }
    \label{fig:sens-teastore}
\end{minipage}
\hfill
\begin{minipage}[t]{0.35\textwidth}
    \centering
    \includegraphics[width=\textwidth]{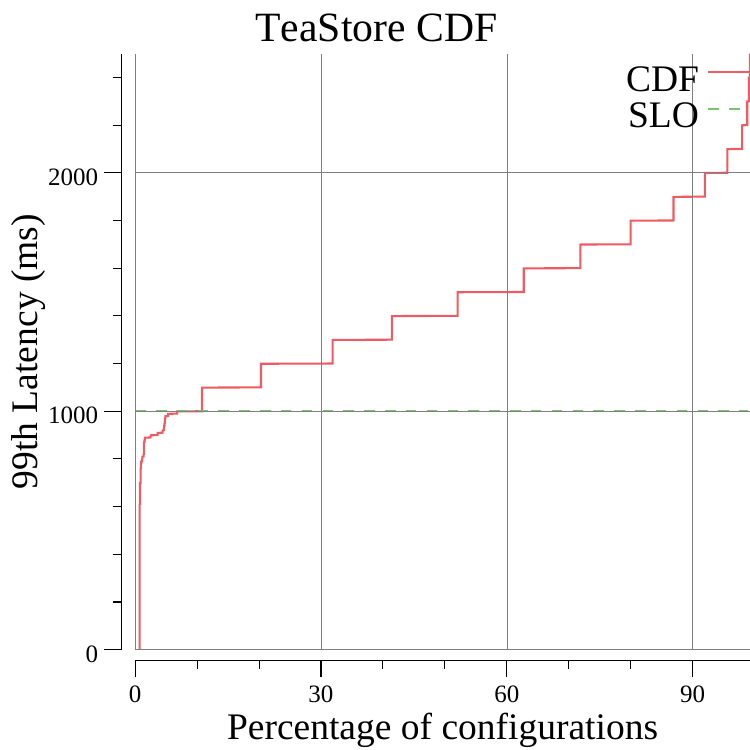}
    \caption{99th percentile latency CDF of exhaustive search space.}
    \label{fig:exhaustive-SLO-cdf}
\end{minipage}
\end{figure}

The results of the factor screening algorithm are listed in Table~\ref{table:sens-tea} and the 99th percentile latency of the samples evaluated during factor screening is plotted in Figure~\ref{fig:sens-teastore}.  Figure~\ref{fig:sens-teastore} shows the clear impact of configuration settings on the observed latency. However, data points are also often clustered indicating that some parameter changes have a minimal effect on the observed latency.   
From Table~\ref{table:sens-tea} it is clear that some parameters have a larger impact on the 99th percentile latency, i.e., those with a higher $\mu^*$ and $\sigma$ values. As will be shown in Section \ref{subsec:resource-utilization} on resource utilization, these parameters also constitute effective performance bottlenecks within the TeaStore application.

As explained in Section ~\ref{sec:meth:subsec:sens}, the new \textit{$min_{i}$} is set to the lowest parameter value of $x_{i}$ achieving a less strict SLO (see Table~\ref{table:sens-tea}). The $max_{i}$ is based on the maximum parameter value of $x_{i}$ that achieves a stricter SLO and the relative impact of $(\mu^*,\sigma)$. These new settings are also shown in Table~\ref{table:sens-tea}.  For TeaStore, the search space is reduced to 2048 configurations, pointing to a enormous reduction factor of $10^7$. 

\begin{table}
\centering
\begin{minipage}[t]{0.35\textwidth}
\vspace{-5cm}
\centering
\fontsize{8}{10}\selectfont
\centering
\begin{tabular}{@{}p{3.5cm}p{3cm}@{}}
\toprule
 & \textbf{TeaStore} \\ \midrule
\textbf{SLO} - \textbf{99th prctl latency} \newline \textbf{relaxed/target/strict (ms)} & 1250/1000/750 \\
\textbf{SLO} - \textbf{workload intensity} \newline \textbf{(\# concurrent tenants)} & 10 \\ 
\textbf{Initial CPU bounds (millicore)} & [500, 1125] \\
\textbf{Initial Mem bounds (Mi)} & [512, 1152] \\
\textbf{Initial search space (configs)} & $6^{14}$ \\
\textbf{Reduced search space (configs)} & 2048 \\
\textbf{Search space reduction} & $10^7$ \\
\textbf{Sample time (min)} & 10 \\
\bottomrule
\end{tabular}
\caption{Initial search spaces, reduced search spaces and sample time of applications.}
\label{table:sens-initial-bounds}
\end{minipage}%
\hfill
\begin{minipage}[t]{0.61\textwidth}
\includegraphics[width=\textwidth]{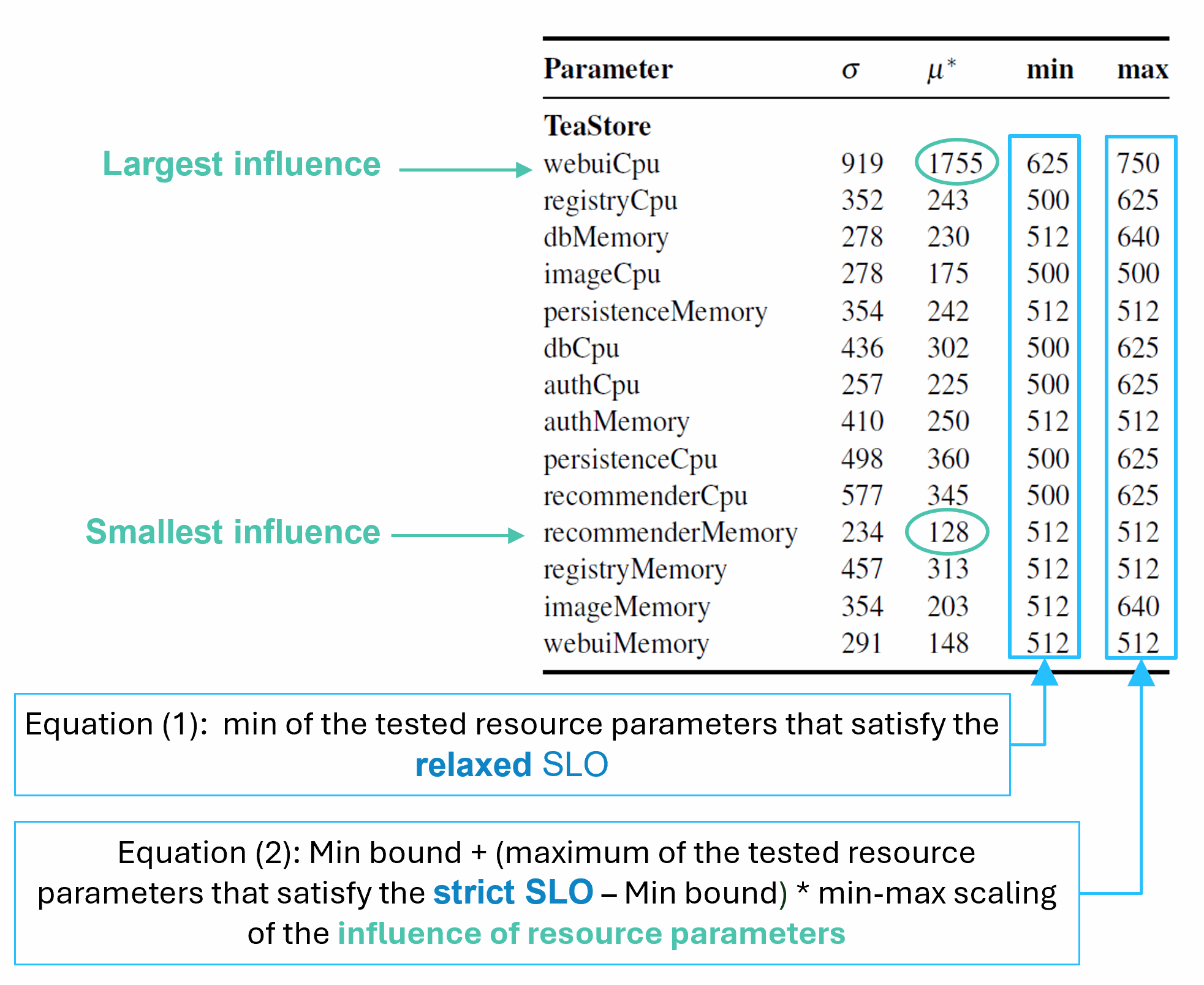}
\caption{Results of the factor screening for TeaStore include the modified mean $\mu^*$ and the deviation $\sigma$. The newly selected [min, max] parameter search spaces are then computed with the formulas shown.}
\label{table:sens-tea}
\end{minipage}
\end{table}

\subsection{Exhaustive data collection}
\label{subsec:eval:exhaustive}
The methodology for comparing the different optimization algorithms entails to exhaustively evaluate all configurations of the reduced search space that resulted from factor screening, and to store the evaluations in a persistent dataset to be leveraged for effective comparison of the optimization algorithms. A similar approach for evaluation of optimization algorithms has been used for optimization of storage solutions~\cite{tuningstorage}. Figure~\ref{fig:exhaustive-SLO-cdf} shows the 99th percentile latency CDF among all configurations for TeaStore. The measured 99th percentile latency values vary across a wide range. Some configurations experience twice the desired SLO. For TeaStore, merely 11\% of the 2048 configurations is capable of satisfying the SLO. This shows that even after factor screening, the search space is still large enough to be explored by the optimization algorithms.

\subsection{Reaching the optimal}
\label{subsec:distance}
As described in Section~\ref{optimizers-explained}, we employ optimization techniques to guide our search in a cost-effective manner (i.e, the number of required performance tests). Here, we focus particularly on 1) BO with expected improvement (EI) 2) BestConfig's optimization algorithm 3) Random Search, which performs random selection 4) Random search with increment (randomInc) which iterates over possible configurations with a randomly selected sequence of parameters (see Section~\ref{subsec:expenv} for more information on their implementation).

\begin{figure}[htbp]
     \centering
     \begin{minipage}[t]{0.40\textwidth}
        \centering
        \includegraphics[width=\textwidth]{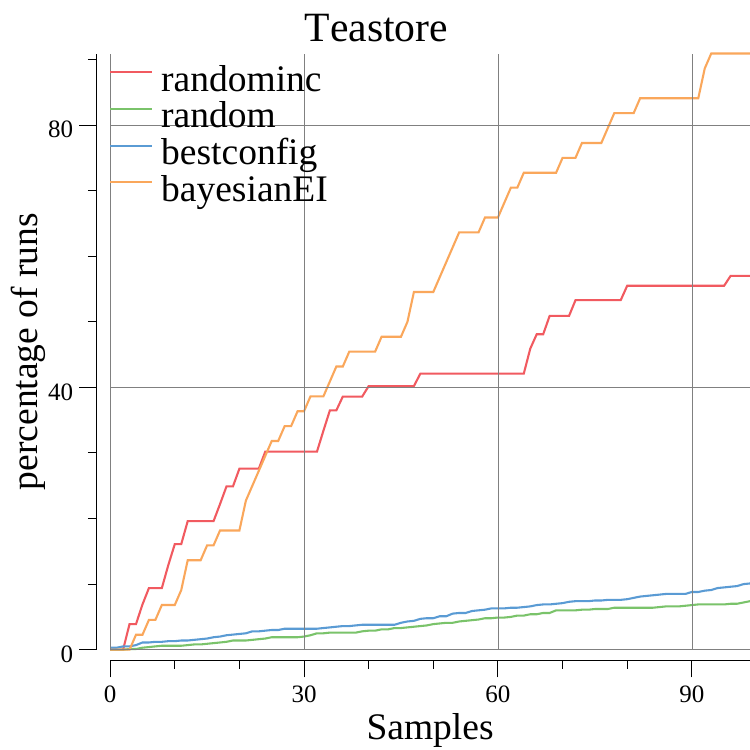}
        \caption{Comparing the efficacy of implemented optimization algorithms in finding the optimal configuration in the reduced search space. The Y-axis shows the percentage of 1,000 runs that found the optimal configuration within a certain amount of a samples (X-axis).}
        \label{fig:exhaustive-cdf}
		 \end{minipage}
			\hfill
			\begin{minipage}[t]{0.40\textwidth}
        \centering
        \includegraphics[width=\textwidth]{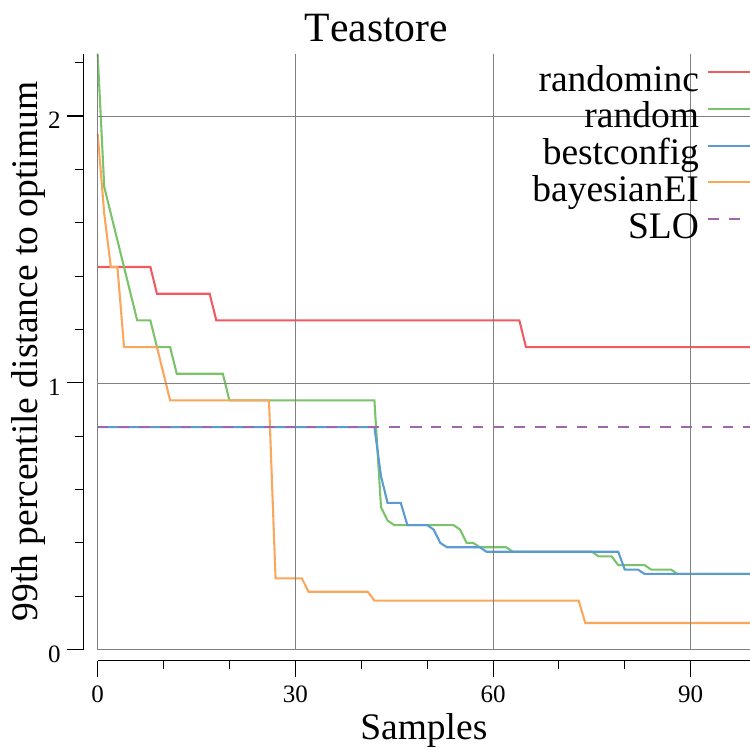}
         \caption{Comparing the worst-case efficacy of optimization methods in finding the near-optimal configurations in the reduced search space. The Y-axis shows the 99th percentile (99\% of the runs find similar or better configurations) of the distance to the optimal configuration.}
        \label{fig:distance-cdf}
				\end{minipage}
\end{figure}

The selected optimization algorithms rely on random exploration to get  good coverage of the search space. To account for this randomness in the evaluation, we collected 1,000 runs for each algorithm. 
Figure~\ref{fig:exhaustive-cdf} shows the results of 1,000 runs on the dataset of each algorithm for TeaStore. The Y-axis indicates the percentage of runs that have found the most-optimal configuration (this value is known from the exhaustive search). The number of samples (or iterations of the algorithm) is limited to 100 for TeaStore.  As such, the algorithms can effectively sample $\approx$5\% of the search space.  For all algorithms, it is clearly shown that the more samples are collected, the more runs find the optimal configuration. However, they differ significantly in their efficacy and speed. For both applications, BO with EI performs the best with more than 80\% of the runs finding the optimal configuration within a limited number of samples.  BestConfig and Random search rarely succeed in finding the optimal solution.

\subsection{Reaching the near-optimal the most quickly}
\label{subsec:eval:comparative}
The previous section evaluated the capacity of the optimization algorithms to find the optimal configuration as found by exhaustive search. In practice, however, near-optimal configurations that satisfy the SLO with reasonable cost reduction are equally desirable. The goal is to offer guidance to select the most effective algorithm given an available sampling cost. In addition, because all algorithms have a random element, it is better to compare the worst performing run of the algorithms.  Figure~\ref{fig:distance-cdf} shows the 99th percentile distance to the most-optimal configuration for the 1,000 simulated runs for each algorithm. The purple dashes indicate the distance between the utility score (1.0) at which the SLO is satisfied (as expressed by the employed utility function) and the optimal configuration as found by exhaustive search.

The results show that BO with EI is the most cost-effective in finding a near-optimal configuration, but BestConfig and random search are also still cost-effective in finding such configurations. However, randomInc does not always find a configuration that meets the SLO because by keeping the order of parameters after shuffling fixed, it may not find such a solution after 100 samples.   

\subsection{Evaluation of factor screening}
\label{subsec:screening}
\begin{figure}
    \centering
    \includegraphics[width=0.60\textwidth]{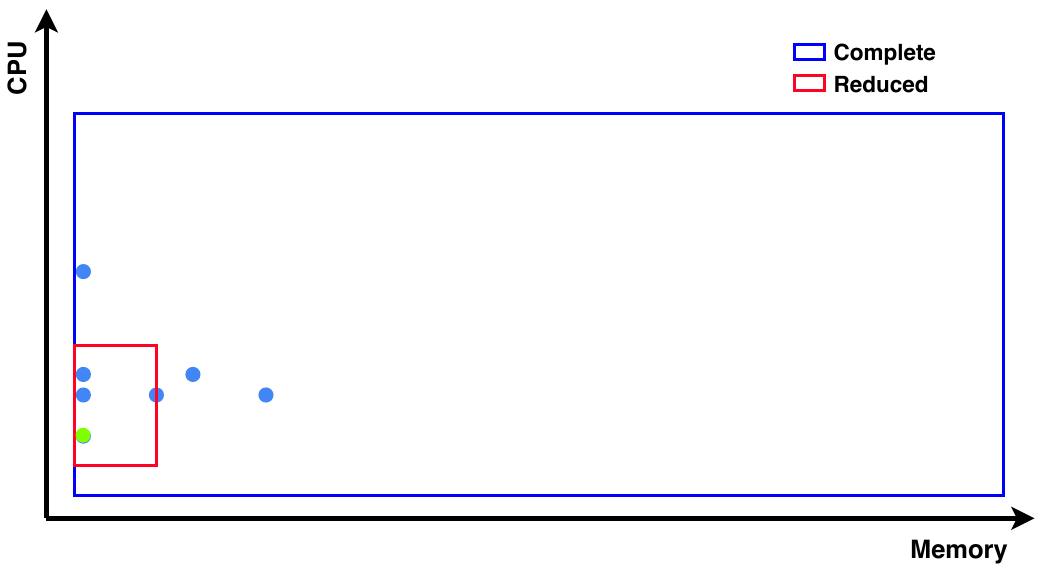}
    \caption{Results of stand-alone BO compared to combining factor screening and BO for TeaStore. The red box represents the reduced search space determined by the factor screening method. The green dot indicates the best result when combining screening and BO. Blue dots are the best found configurations by BO.}
    \label{fig:exhaustive-vs-complete}
\end{figure}

Factor creening is set at 150 samples. This is an significant cost. The question arises whether applying the algorithms on the full parameter space without screening would have resulted in finding the optimal configuration faster or even finding a more optimal configuration than found by the exhaustive search within the bounds set of the screening algorithm. To answer this question, we ran the BO algorithm at 150 samples on the complete search space for TeaStore before screening (cfr. the initial bounds set in Table \ref{table:sens-initial-bounds}). Figure \ref{fig:exhaustive-vs-complete} shows the results. The red box represents the search space selected by the screening method, with the most expensive configuration in the top red corner. The red dot shows the best result found when combining factor screening and BO, the blue dots indicate the best-found configurations by running BO on the complete search space without factor screening. This shows that standalone BO can find near-optimal solutions as well using the same cost as the screening cost. Thus it can find near-optimal solutions faster than when combining BO and screening, which supports more efficient performance optimization during the Release phase of the DevOps lifecycle, where rapid and accurate resource tuning is critical.

However, combining factor screening and BO hugely increases the probability to find the optimal configuration, while this is not the case when running BO stand-alone. As such we conclude that factor screening remain advisable for finding the optimal configuration without caring for the sampling cost. 

Of course, factor screening is also required for reducing the effort to statistically compare multiple optimization algorithms (cfr. Section \ref{subsec:distance}). This enables fast, automated comparison of optimization strategies, which aligns with the need for efficient tooling in DevOps. Indeed, factor screening drastically reduces the search space, making it feasible to collect samples for all data points in limited amount of time, which is valuable during the Release phase where time constraints often limit the feasibility of full search strategies. Thereafter each algorithm can be evaluated very quickly by running k8-resource-optimizer from the collected dataset multiple times. Repeating 1000 runs takes about 6 minutes on average. Still, sequentially collecting all 2048 samples took more than 14 days. We also applied factor screening for TeaStore version 1.3.0 and this reduced the search space to 128 samples, but still requiring a full day (see Table \ref{table:duration} for a break-down of the durations).

\begin{table}[h!]
 \fontsize{8}{10}\selectfont
\centering
\begin{tabular}{@{}lllll@{}}
\toprule
 & \textbf{Sample} & \textbf{Screening (150 samples)} & \textbf{Exhaustive} & \textbf{1000 optimization runs} \\ \midrule
\textbf{TeaStore  v1.2} & 10 min & 25 hours & 14.2 days & 7 min \\
\textbf{TeaStore v1.3} & 10 min & 25 hours & 21.5 hours & 5 min \\
\end{tabular}
\caption{Break down duration cost without parallel execution. }
\label{table:duration}
\end{table}

\subsection{Resource utilization}
\label{subsec:resource-utilization}
\begin{figure}[htbp]
     \centering
      \includegraphics[width=0.40\textwidth]{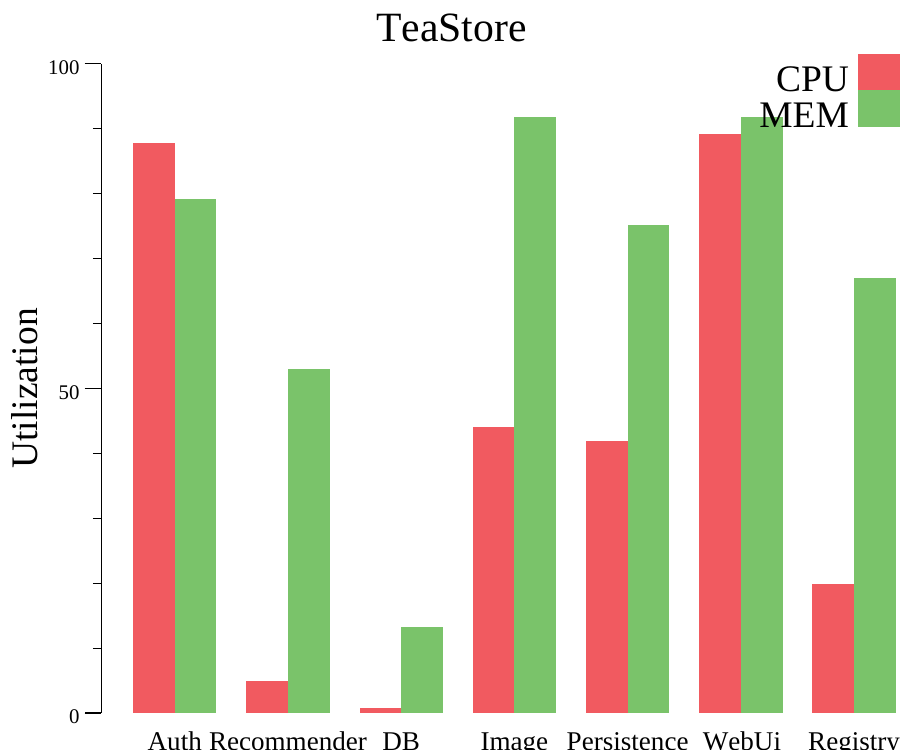}
      \caption{Worst case utilization of separate microservices after optimization. The resource parameters that have been determined as most influential by factor screening appear as performance bottlenecks. Note that low CPU utilization is due to high Java startup overhead. Future work aims to integrate k8-resource-optimizer with on-line resizing of containers without restart \cite{truyen-woc2020}}
        \label{fig:utilization}
\end{figure}

Figure~\ref{fig:utilization} illustrates the resource utilization of the resource configurations that have been found by the worst-case run of the most cost-effective algorithm after screening and after the maximum amount of samples. This resource utilization has been measured using K8s' Heapster monitoring service~\cite{CloudNativeComputingFoundationHeapster}.
In the case of TeaStore, a mix of high and low utilization of resources between the different services is observed. The reasons for this is twofold: firstly, the application's bottlenecks are located at the heavily utilized Authentication and WebUI service.  Secondly, other services such as the Recommendation, Registry and Database services experience a low degree of utilization in terms of CPU despite although the configuration selected the lowest possible CPU setting for these settings. This is because a higher amount of CPU is needed for starting up the Java-based services, but once the microservices are started up, the CPU usage tends to be lower. Unfortunately, at the time of the research, in K8s v1.14, it is not possible to resize the resources of containers without restarting these containers, which limits flexibility during the Release phase, when rapid adjustments with minimal disruption are crucial. But this feature is now in alpha development \cite{truyen-woc2020}. K8s-resource optimizer could therefore be extended so that two CPU parameters are optimized, one for starting the containers and when one warmed up and processing workload. 

Setting minimal bounds is a tricky problem as a too high minimal bound may exclude the optimal configuration. This is illustrated by the low memory usage of the database component. We made the wrong assumption that the database service also ran on Tomcat but it does not. Therefore, instead of setting the same minimal bound for all microservices (cfr. Table \ref{table:sens-initial-bounds}), we should have tested the minimal bound for each microservice separately.

\section{Towards optimization of adaptation-relevant API and application parameters}
\label{discussion}

The evaluation results in Sections~\ref{subsec:screening} and~\ref{subsec:resource-utilization} demonstrate the practicality of using k8-resource-optimizer during the Release phase of the DevOps lifecycle to discover cost-efficient resource configurations for attaining performance SLOs of microservices. While our evaluation focuses on resource-level configuration, the same methodology directly applies to parameters that govern adaptive behavior of microservices -- an increasingly important concern in systems such as Adaptable TeaStore\cite{BDGLZZ25}. A key takeaway is that standalone BO can be highly effective in identifying near-optimal configurations with the same sampling cost as the screening algorithm. However, when the objective shifts to finding the optimal configuration with higher certainty, or enabling statistical comparisons across optimizers, factor screening remains essential despite its upfront cost.

As discussed in Section~\ref{introduction}, k8-resource-optimizer is not restricted to container-level resource parameters but any parameter that is configurable via the K8s API. As shown in Appendix \ref{app:string-param}, such application-level parameters can be included in the optimizer’s search space via Helm chart templating. By representing them as variables in the \texttt{values.yaml} file, developers can explore a much broader configuration space. This enables optimization not only of infrastructure-level settings, but also of adaptive functional behaviors (e.g., selecting alternative execution modes, enabling/disabling features, or adjusting quality levels), which are central mechanisms for adaptability in modern microservice architectures.

Application-specific configurations of microservices can thus also be included in the search space. Examples include adaptation-oriented feature toggles, such as switching from high-power to low-power mode under resource contention. Through optimization runs in the Release phase, k8-resource-optimizer can identify performance invariants that guide runtime adaptation policies. For example, the following performance invariant can be directly obtained from k8-resource optimizer: \textit{high power mode of a microservice only attains the desired SLO when CPU allocation is higher than 750 millicores}. Such an invariant can inform adaptation logic -- for example, a controller could enter low-power mode below 750 millicores or request additional CPU when operating in high-power mode. Thus, optimization results obtained offline can directly shape runtime adaptation strategies. 

However, this increased adaptability poses challenges: as more adaptation knobs are exposed, the configuration space grows and interactions become harder to reason about. Screening therefore becomes a key enabler for practical and scalable adaptation-policy design. The inclusion of application-specific parameters increases the dimensionality of the search space, making techniques like factor screening more valuable for pruning irrelevant parameters. Moreover, interactions between resource settings and application configurations may create complex dependencies that require joint consideration during optimization.

\section{Related work}
\label{related_work}
This section presents existing work in the area of performance optimization during the development stage of the DevOps life cycle. We refer the reader for an overview of performance optimization during the operational stage to existing surveys \cite{survey_buyya_2022,moreschini2025ai}.

\noindent\textbf{Performance modeling.} Previous work has proposed several methods based on performance modeling of microservice applications \cite{9215019, performance_modeling}. The performance model estimates the resources required for an SLO or the maximum request rate without violating the SLO. Performance models can either be constructed using queuing theory~\cite{chen2007sla} or via various supervised machine learning techniques such as neural networks or state vector machines~\cite{kundu2012modeling}.  The quality of the queuing model or machine learning model relies heavily on the expertise of the application development team. In opposition, extending k8-resource optimizer for another microservice application requires implementing an end-to-end benchmark scenario, an implementation of an appropriate utility function, and the definition of an optimizer configuration file that is based on the \texttt{values.yaml} file of the helm chart of the application. The sampling cost of 150 performance tests for factor screening or optimization is in line with the profiling/training cost of the performance modeling approaches. K8-resource optimizer also allows comparing different optimization algorithms with respect to finding the optimal configuration or reaching a near-optimal configuration fast enough. 

\noindent\textbf{Performance optimization for best VM instance selection.}
The closed work to ours are studies on the selection of VM instances to achieve performance SLOs while minimizing costs. \textit{Ernest}~\cite{venkataraman2016ernest} can select VM sizes within a given instance family for various machine learning applications by training a common performance model with a small number of samples. The internal performance model exploits known patterns in ML applications making it poorly adaptive for other types of applications. 
\textit{CherryPick}~\cite{alipourfard2017cherrypick} utilizes Naive BO to find optimal or near-optimal VM instances for recurring big data analytics jobs. \textit{Arrow}~\cite{hsu2018arrow} 
introduces Augmented BO which modifies off-the-shelf BO by integrating low-level performance information (e.g., CPU utilization or work memory allocation) and design choices, allowing for more informed decision making. \textit{Scout}~\cite{hsu2018scout} leverages historical data from the optimization process and low-level performance metrics resulting in a more efficient search. \textit{Micky}~\cite{hsu2018micky} bundles several of the above techniques in a collective optimizer to further reduce measurement costs.

There are clear analogies between k8-resource-optimizer and selecting the cost-optimal VM-instance for an application. Our work deals however with a more complex configuration search space.  Moreover k8-resource-optimizer can be easily reused with different optimization algorithms beyond BO. In opposition, the above works have only focused on BO or very similar search strategies.

\noindent\textbf{Performance optimization of program configuration parameters} is an active research domain. \textit{BestConfig}~\cite{zhu2017bestconfig} tunes general systems with high-dimensional parameter spaces using a recursive search with stratified sampling. Similarly, Latin Hypercube Sampling (LHS) and smart hill climbing have been used to tune an application server~\cite{xi2004smart}. BOAT~\cite{dalibard2017boat}  utilizes structured BO while leveraging contextual information to automatically tune application performance.  The recent work of \textit{Metis}~\cite{li2018metis} enhances BO and demonstrates its capability to tuning the tail latency of the Bing Ads key-value store. 
\textit{AdaptiveConfig}~\cite{ChenLydia} combines an efficient search algorithm and a business rule engine to search at run-time for the optimal job scheduler configuration.
Einziger et al.~\cite{Friedman} compares hill climbing and an indicator-based approach for optimization of adaptive cache management. This work discusses the configuration of the optimization algorithms such as the distance between different samples and the frequency of taking samples. Their findings show that these particular optimization algorithms do not perform differently from each other. 

The main difference between k8-resource-optimizer and the above works is that the latter mainly focuses on optimizing performance, whereas we focus on optimizing performance and resource cost. 

\section{Conclusion}
\label{conclusion}
This paper has presented how to apply k8-resource-optimizer to performance optimization of containerized, microservice applications during the Release phase of the DevOps lifecyle.  When the goal is to find a near-optimal configuration using a limited sampling budget, our findings show that is better to run k8-resource optimizer with bayesian optimization without the use of factor screening/sensitivity analysis. When the goal is the find a configuration that is close to the optimal solution, upfront screening to reduce the search space is advisable. Screening is also required for the statistical comparison of different optimization algorithms so that the collection of a relevant dataset of samples becomes feasible. 
 
The primary limitation of k8-resource-optimizer is that the benchmark scenario used for performance testing is fixed. For example, consider a benchmark involving two \texttt{Experiment} classes that each trigger different API operations of the TeaStore application, with an initial workload distribution of 50\% each. One \texttt{Experiment} class is primarily memory-intensive, while the other is predominantly CPU-intensive. If the overall workload intensity remains constant but the workload distribution shifts to 90\%–10\%, the previously determined resource configuration becomes significantly misaligned with actual demands — resulting in severe under-provisioning for one resource type and over-provisioning for the other. A promising direction for future work is therefore to develop efficient methods for evaluating a mix of different \texttt{Experiment} classes with a range of different workload distributions, thereby accounting for fluctuations in user behavior and enabling interpolation of resource allocations for untested workload distributions. 

Another direction of future work is accounting for multiple versions of the software of the microservices themselves. Consider the deployment of a new version of the TeaStore microservice using a new container image, or a configuration change in the software. Currently, each software version would require a new run of k8-resource optimizer from scratch.  

\section*{Acknowledgments}
This research is partially funded by the Research Fund KU Leuven. We thank the anonymous reviewers of the WACA 2025 PC for their helpful comments and suggestions.

\bibliographystyle{eptcs}

\bibliography{bv2.bib} 

\begin{thebibliography}{10}
\providecommand{\bibitemdeclare}[2]{}
\providecommand{\surnamestart}{}
\providecommand{\surnameend}{}
\providecommand{\urlprefix}{Available at }
\providecommand{\url}[1]{\texttt{#1}}
\providecommand{\href}[2]{\texttt{#2}}
\providecommand{\urlalt}[2]{\href{#1}{#2}}
\providecommand{\doi}[1]{doi:\urlalt{https://doi.org/#1}{#1}}
\providecommand{\eprint}[1]{arXiv:\urlalt{https://arxiv.org/abs/#1}{#1}}
\providecommand{\bibinfo}[2]{#2}

\bibitemdeclare{inproceedings}{alipourfard2017cherrypick}
\bibitem{alipourfard2017cherrypick}
\bibinfo{author}{Omid \surnamestart Alipourfard\surnameend},
  \bibinfo{author}{Hongqiang~Harry \surnamestart Liu\surnameend},
  \bibinfo{author}{Jianshu \surnamestart Chen\surnameend},
  \bibinfo{author}{Shivaram \surnamestart Venkataraman\surnameend},
  \bibinfo{author}{Minlan \surnamestart Yu\surnameend} \& \bibinfo{author}{Ming
  \surnamestart Zhang\surnameend} (\bibinfo{year}{2017}):
  \emph{\bibinfo{title}{CherryPick: Adaptively Unearthing the Best Cloud
  Configurations for Big Data Analytics.}}
\newblock In: {\slshape \bibinfo{booktitle}{NSDI}}, \bibinfo{volume}{2}, pp.
  \bibinfo{pages}{4--2}.
\newblock
  \urlprefix\url{https://www.usenix.org/conference/nsdi17/technical-sessions/presentation/alipourfard}.

\bibitemdeclare{article}{balalaie2016microservices}
\bibitem{balalaie2016microservices}
\bibinfo{author}{Armin \surnamestart Balalaie\surnameend},
  \bibinfo{author}{Abbas \surnamestart Heydarnoori\surnameend} \&
  \bibinfo{author}{Pooyan \surnamestart Jamshidi\surnameend}
  (\bibinfo{year}{2016}): \emph{\bibinfo{title}{Microservices architecture
  enables devops: Migration to a cloud-native architecture}}.
\newblock {\slshape \bibinfo{journal}{IEEE Software}}
  \bibinfo{volume}{33}(\bibinfo{number}{3}), pp. \bibinfo{pages}{42--52},
  \doi{10.1109/MS.2016.64}.

\bibitemdeclare{book}{beyer2018site}
\bibitem{beyer2018site}
\bibinfo{author}{Betsy \surnamestart Beyer\surnameend},
  \bibinfo{author}{Niall~Richard \surnamestart Murphy\surnameend},
  \bibinfo{author}{David~K \surnamestart Rensin\surnameend},
  \bibinfo{author}{Kent \surnamestart Kawahara\surnameend} \&
  \bibinfo{author}{Stephen \surnamestart Thorne\surnameend}
  (\bibinfo{year}{2018}): \emph{\bibinfo{title}{The Site Reliability Workbook:
  Practical Ways to Implement SRE}}.
\newblock \bibinfo{publisher}{" O'Reilly Media, Inc."}.
\newblock
  \urlprefix\url{https://lrita.github.io/images/posts/com/the-site-reliability-workbook-next18.pdf}.

\bibitemdeclare{inproceedings}{BDGLZZ25}
\bibitem{BDGLZZ25}
\bibinfo{author}{Simon \surnamestart Bliudze\surnameend},
  \bibinfo{author}{Giuseppe \surnamestart {De Palma}\surnameend},
  \bibinfo{author}{Saverio \surnamestart Giallorenzo\surnameend},
  \bibinfo{author}{Ivan \surnamestart Lanese\surnameend},
  \bibinfo{author}{Gianluigi \surnamestart Zavattaro\surnameend} \&
  \bibinfo{author}{Brice~Arl{\'{e}}on \surnamestart {Zemtsop
  Ndadji}\surnameend} (\bibinfo{year}{2025}): \emph{\bibinfo{title}{Adaptable
  TeaStore}}.
\newblock In \bibinfo{editor}{Giuseppe \surnamestart {De Palma}\surnameend} \&
  \bibinfo{editor}{Saverio \surnamestart Giallorenzo\surnameend}, editors:
  {\slshape \bibinfo{booktitle}{Post-proceedings of the Workshop on Adaptable
  Cloud Architectures (WACA 2025)}}, \bibinfo{series}{\thisvolume{9}}.

\bibitemdeclare{inproceedings}{tuningstorage}
\bibitem{tuningstorage}
\bibinfo{author}{Zhen \surnamestart Cao\surnameend}, \bibinfo{author}{Sachin
  \surnamestart Tiwari\surnameend}, \bibinfo{author}{Erez \surnamestart
  Zadok\surnameend} \& \bibinfo{author}{Vasily \surnamestart
  Tarasov\surnameend} (\bibinfo{year}{2018}): \emph{\bibinfo{title}{{Towards
  Better Understanding of Black-box Auto-Tuning: A Comparative Analysis for
  Storage Systems}}}.
\newblock In: {\slshape \bibinfo{booktitle}{USENIX Annual Technical
  Conference}}, pp. \bibinfo{pages}{893--907}.
\newblock
  \urlprefix\url{https://www.usenix.org/conference/atc18/presentation/cao}.

\bibitemdeclare{inproceedings}{chen2007sla}
\bibitem{chen2007sla}
\bibinfo{author}{Yuan \surnamestart Chen\surnameend}, \bibinfo{author}{Subu
  \surnamestart Iyer\surnameend}, \bibinfo{author}{Xue \surnamestart
  Liu\surnameend}, \bibinfo{author}{Dejan \surnamestart Milojicic\surnameend}
  \& \bibinfo{author}{Akhil \surnamestart Sahai\surnameend}
  (\bibinfo{year}{2007}): \emph{\bibinfo{title}{{SLA decomposition: Translating
  service level objectives to system level thresholds}}}.
\newblock In: {\slshape \bibinfo{booktitle}{Autonomic Computing, 2007. ICAC'07.
  Fourth International Conference on}}, \bibinfo{organization}{IEEE}, pp.
  \bibinfo{pages}{3--3}, \doi{10.1109/ICAC.2007.36}.

\bibitemdeclare{misc}{CloudNativeComputingFoundation}
\bibitem{CloudNativeComputingFoundation}
\bibinfo{author}{\surnamestart {Cloud Native Computing Foundation}\surnameend}
  (\bibinfo{year}{2019}): \emph{\bibinfo{title}{{Configure Quality of Service
  for Pods - Kubernetes}}}.
\newblock
  \urlprefix\url{https://kubernetes.io/docs/tasks/configure-pod-container/quality-service-pod/}.

\bibitemdeclare{misc}{CloudNativeComputingFoundationHeapster}
\bibitem{CloudNativeComputingFoundationHeapster}
\bibinfo{author}{\surnamestart {Cloud Native Computing Foundation}\surnameend}
  (\bibinfo{year}{2019}): \emph{\bibinfo{title}{{Tools for Monitoring Compute,
  Storage, and Network Resources - Kubernetes}}}.
\newblock
  \urlprefix\url{https://kubernetes.io/docs/tasks/debug-application-cluster/resource-usage-monitoring/}.

\bibitemdeclare{inproceedings}{dalibard2017boat}
\bibitem{dalibard2017boat}
\bibinfo{author}{Valentin \surnamestart Dalibard\surnameend},
  \bibinfo{author}{Michael \surnamestart Schaarschmidt\surnameend} \&
  \bibinfo{author}{Eiko \surnamestart Yoneki\surnameend}
  (\bibinfo{year}{2017}): \emph{\bibinfo{title}{BOAT: Building auto-tuners with
  structured Bayesian optimization}}.
\newblock In: {\slshape \bibinfo{booktitle}{Proceedings of the 26th
  International Conference on World Wide Web}},
  \bibinfo{organization}{International World Wide Web Conferences Steering
  Committee}, pp. \bibinfo{pages}{479--488}, \doi{10.1145/3038912.3052662}.

\bibitemdeclare{inproceedings}{Friedman}
\bibitem{Friedman}
\bibinfo{author}{Gil \surnamestart Einziger\surnameend}, \bibinfo{author}{Ohad
  \surnamestart Eytan\surnameend}, \bibinfo{author}{Roy \surnamestart
  Friedman\surnameend} \& \bibinfo{author}{Ben \surnamestart Manes\surnameend}
  (\bibinfo{year}{2018}): \emph{\bibinfo{title}{Adaptive Software Cache
  Management}}.
\newblock In: {\slshape \bibinfo{booktitle}{Proceedings of the 19th
  International Middleware Conference}}, \bibinfo{series}{Middleware '18},
  \bibinfo{publisher}{ACM}, \bibinfo{address}{New York, NY, USA}, pp.
  \bibinfo{pages}{94--106}, \doi{10.1145/3274808.3274816}.

\bibitemdeclare{inproceedings}{ChenLydia}
\bibitem{ChenLydia}
\bibinfo{author}{R.~\surnamestart {Han}\surnameend},
  \bibinfo{author}{Z.~\surnamestart {Zong}\surnameend}, \bibinfo{author}{L.~Y.
  \surnamestart {Chen}\surnameend}, \bibinfo{author}{S.~\surnamestart
  {Wang}\surnameend} \& \bibinfo{author}{J.~\surnamestart {Zhan}\surnameend}
  (\bibinfo{year}{2018}): \emph{\bibinfo{title}{AdaptiveConfig: Run-Time
  Configuration of Cluster Schedulers for Cloud Short-Running Jobs}}.
\newblock In: {\slshape \bibinfo{booktitle}{2018 IEEE 38th International
  Conference on Distributed Computing Systems (ICDCS)}}, pp.
  \bibinfo{pages}{1519--1526}, \doi{10.1109/ICDCS.2018.00158}.

\bibitemdeclare{misc}{LocustHeyman}
\bibitem{LocustHeyman}
\bibinfo{author}{Jonatan \surnamestart Heyman\surnameend},
  \bibinfo{author}{Carl \surnamestart Bystr{\"{o}}m\surnameend},
  \bibinfo{author}{Joakim \surnamestart Hamr{\'{e}}n\surnameend} \&
  \bibinfo{author}{Hugo \surnamestart Heyman\surnameend}
  (\bibinfo{year}{2019}): \emph{\bibinfo{title}{{Locust - A modern load testing
  framework}}}.
\newblock \urlprefix\url{https://locust.io/}.

\bibitemdeclare{inproceedings}{hsu2018arrow}
\bibitem{hsu2018arrow}
\bibinfo{author}{Chin-Jung \surnamestart Hsu\surnameend},
  \bibinfo{author}{Vivek \surnamestart Nair\surnameend},
  \bibinfo{author}{Vincent~W \surnamestart Freeh\surnameend} \&
  \bibinfo{author}{Tim \surnamestart Menzies\surnameend}
  (\bibinfo{year}{2018}): \emph{\bibinfo{title}{Arrow: Low-Level Augmented
  Bayesian Optimization for Finding the Best Cloud VM}}.
\newblock In: {\slshape \bibinfo{booktitle}{2018 IEEE 38th International
  Conference on Distributed Computing Systems (ICDCS)}},
  \bibinfo{organization}{IEEE}, pp. \bibinfo{pages}{660--670},
  \doi{10.1109/ICDCS.2018.00070}.

\bibitemdeclare{article}{hsu2018micky}
\bibitem{hsu2018micky}
\bibinfo{author}{Chin-Jung \surnamestart Hsu\surnameend},
  \bibinfo{author}{Vivek \surnamestart Nair\surnameend}, \bibinfo{author}{Tim
  \surnamestart Menzies\surnameend} \& \bibinfo{author}{Vincent \surnamestart
  Freeh\surnameend} (\bibinfo{year}{2018}): \emph{\bibinfo{title}{Micky: A
  Cheaper Alternative for Selecting Cloud Instances}}.
\newblock {\slshape \bibinfo{journal}{arXiv preprint arXiv:1803.05587}},
  \doi{10.48550/arXiv.1803.05587}.

\bibitemdeclare{article}{hsu2018scout}
\bibitem{hsu2018scout}
\bibinfo{author}{Chin-Jung \surnamestart Hsu\surnameend},
  \bibinfo{author}{Vivek \surnamestart Nair\surnameend}, \bibinfo{author}{Tim
  \surnamestart Menzies\surnameend} \& \bibinfo{author}{Vincent~W \surnamestart
  Freeh\surnameend} (\bibinfo{year}{2018}): \emph{\bibinfo{title}{Scout: An
  Experienced Guide to Find the Best Cloud Configuration}}.
\newblock {\slshape \bibinfo{journal}{arXiv preprint arXiv:1803.01296}},
  \doi{10.48550/arXiv.1803.01296}.

\bibitemdeclare{inproceedings}{performance_modeling}
\bibitem{performance_modeling}
\bibinfo{author}{Anshul \surnamestart Jindal\surnameend},
  \bibinfo{author}{Vladimir \surnamestart Podolskiy\surnameend} \&
  \bibinfo{author}{Michael \surnamestart Gerndt\surnameend}
  (\bibinfo{year}{2019}): \emph{\bibinfo{title}{Performance Modeling for Cloud
  Microservice Applications}}.
\newblock In: {\slshape \bibinfo{booktitle}{Proceedings of the 2019 ACM/SPEC
  International Conference on Performance Engineering}}, \bibinfo{series}{ICPE
  '19}, \bibinfo{publisher}{Association for Computing Machinery}, pp.
  \bibinfo{pages}{25--32}, \doi{10.1145/3297663.3310309}.

\bibitemdeclare{inproceedings}{jyothi2016morpheus}
\bibitem{jyothi2016morpheus}
\bibinfo{author}{Sangeetha~Abdu \surnamestart Jyothi\surnameend},
  \bibinfo{author}{Carlo \surnamestart Curino\surnameend},
  \bibinfo{author}{Ishai \surnamestart Menache\surnameend},
  \bibinfo{author}{Shravan~Matthur \surnamestart Narayanamurthy\surnameend},
  \bibinfo{author}{Alexey \surnamestart Tumanov\surnameend},
  \bibinfo{author}{Jonathan \surnamestart Yaniv\surnameend},
  \bibinfo{author}{Ruslan \surnamestart Mavlyutov\surnameend},
  \bibinfo{author}{Inigo \surnamestart Goiri\surnameend},
  \bibinfo{author}{Subru \surnamestart Krishnan\surnameend},
  \bibinfo{author}{Janardhan \surnamestart Kulkarni\surnameend} et~al.
  (\bibinfo{year}{2016}): \emph{\bibinfo{title}{Morpheus: Towards Automated
  SLOs for Enterprise Clusters.}}
\newblock In: {\slshape \bibinfo{booktitle}{OSDI}}, pp.
  \bibinfo{pages}{117--134}.
\newblock
  \urlprefix\url{https://www.usenix.org/conference/osdi16/technical-sessions/presentation/jyothi}.

\bibitemdeclare{inproceedings}{kaminskiwoc2019}
\bibitem{kaminskiwoc2019}
\bibinfo{author}{Matthijs \surnamestart Kaminski\surnameend},
  \bibinfo{author}{Eddy \surnamestart Truyen\surnameend},
  \bibinfo{author}{Emad~Heydari \surnamestart Beni\surnameend},
  \bibinfo{author}{Bert \surnamestart Lagaisse\surnameend} \&
  \bibinfo{author}{Wouter \surnamestart Joosen\surnameend}
  (\bibinfo{year}{2019}): \emph{\bibinfo{title}{A framework for black-box SLO
  tuning of multi-tenant applications in Kubernetes}}.
\newblock In: {\slshape \bibinfo{booktitle}{Proceedings of the 5th
  International Workshop on Container Technologies and Container Clouds}},
  \bibinfo{series}{WOC '19}, \bibinfo{publisher}{Association for Computing
  Machinery}, \bibinfo{address}{New York, NY, USA}, pp. \bibinfo{pages}{7--12},
  \doi{10.1145/3366615.3368352}.

\bibitemdeclare{article}{9215019}
\bibitem{9215019}
\bibinfo{author}{Hamzeh \surnamestart Khazaei\surnameend},
  \bibinfo{author}{Nima \surnamestart Mahmoudi\surnameend},
  \bibinfo{author}{Cornel \surnamestart Barna\surnameend} \&
  \bibinfo{author}{Marin \surnamestart Litoiu\surnameend}
  (\bibinfo{year}{2022}): \emph{\bibinfo{title}{Performance Modeling of
  Microservice Platforms}}.
\newblock {\slshape \bibinfo{journal}{IEEE Transactions on Cloud Computing}}
  \bibinfo{volume}{10}(\bibinfo{number}{4}), pp. \bibinfo{pages}{2848--2862},
  \doi{10.1109/TCC.2020.3029092}.

\bibitemdeclare{inproceedings}{von2018teastore}
\bibitem{von2018teastore}
\bibinfo{author}{J{\'o}akim \surnamestart von Kistowski\surnameend},
  \bibinfo{author}{Simon \surnamestart Eismann\surnameend},
  \bibinfo{author}{Norbert \surnamestart Schmitt\surnameend},
  \bibinfo{author}{Andr{\'e} \surnamestart Bauer\surnameend},
  \bibinfo{author}{Johannes \surnamestart Grohmann\surnameend} \&
  \bibinfo{author}{Samuel \surnamestart Kounev\surnameend}
  (\bibinfo{year}{2018}): \emph{\bibinfo{title}{TeaStore: A Micro-Service
  Reference Application for Benchmarking, Modeling and Resource Management
  Research}}.
\newblock In: {\slshape \bibinfo{booktitle}{2018 IEEE 26th International
  Symposium on Modeling, Analysis, and Simulation of Computer and
  Telecommunication Systems (MASCOTS)}}, \bibinfo{organization}{IEEE}, pp.
  \bibinfo{pages}{223--236}, \doi{10.1109/MASCOTS.2018.00030}.

\bibitemdeclare{article}{kratzke2017understanding}
\bibitem{kratzke2017understanding}
\bibinfo{author}{Nane \surnamestart Kratzke\surnameend} \&
  \bibinfo{author}{Peter-Christian \surnamestart Quint\surnameend}
  (\bibinfo{year}{2017}): \emph{\bibinfo{title}{Understanding cloud-native
  applications after 10 years of cloud computing-a systematic mapping study}}.
\newblock {\slshape \bibinfo{journal}{Journal of Systems and Software}}
  \bibinfo{volume}{126}, pp. \bibinfo{pages}{1--16},
  \doi{10.1016/j.jss.2017.01.001}.

\bibitemdeclare{inproceedings}{kundu2012modeling}
\bibitem{kundu2012modeling}
\bibinfo{author}{Sajib \surnamestart Kundu\surnameend}, \bibinfo{author}{Raju
  \surnamestart Rangaswami\surnameend}, \bibinfo{author}{Ajay \surnamestart
  Gulati\surnameend}, \bibinfo{author}{Ming \surnamestart Zhao\surnameend} \&
  \bibinfo{author}{Kaushik \surnamestart Dutta\surnameend}
  (\bibinfo{year}{2012}): \emph{\bibinfo{title}{Modeling virtualized
  applications using machine learning techniques}}.
\newblock In: {\slshape \bibinfo{booktitle}{ACM Sigplan Notices}},
  \bibinfo{volume}{47}, \bibinfo{organization}{ACM}, pp.
  \bibinfo{pages}{3--14}, \doi{10.1145/2365864.2151028}.

\bibitemdeclare{inproceedings}{lahmann2018container}
\bibitem{lahmann2018container}
\bibinfo{author}{Garrett \surnamestart Lahmann\surnameend},
  \bibinfo{author}{Thom \surnamestart McCann\surnameend} \&
  \bibinfo{author}{Wes \surnamestart Lloyd\surnameend} (\bibinfo{year}{2018}):
  \emph{\bibinfo{title}{Container Memory Allocation Discrepancies: An
  Investigation on Memory Utilization Gaps for Container-Based Application
  Deployments}}.
\newblock In: {\slshape \bibinfo{booktitle}{Cloud Engineering (IC2E), 2018 IEEE
  International Conference on}}, \bibinfo{organization}{IEEE}, pp.
  \bibinfo{pages}{404--405}, \doi{10.1109/IC2E.2018.00076}.

\bibitemdeclare{inproceedings}{li2018metis}
\bibitem{li2018metis}
\bibinfo{author}{Zhao~Lucis \surnamestart Li\surnameend},
  \bibinfo{author}{Chieh-Jan~Mike \surnamestart Liang\surnameend},
  \bibinfo{author}{Wenjia \surnamestart He\surnameend},
  \bibinfo{author}{Lianjie \surnamestart Zhu\surnameend},
  \bibinfo{author}{Wenjun \surnamestart Dai\surnameend}, \bibinfo{author}{Jin
  \surnamestart Jiang\surnameend} \& \bibinfo{author}{Guangzhong \surnamestart
  Sun\surnameend} (\bibinfo{year}{2018}): \emph{\bibinfo{title}{Metis: Robustly
  Tuning Tail Latencies of Cloud Systems}}.
\newblock In: {\slshape \bibinfo{booktitle}{2018 USENIX Annual Technical
  Conference (USENIX ATC 18)}}, pp. \bibinfo{pages}{981--992}.
\newblock
  \urlprefix\url{https://www.usenix.org/conference/atc18/presentation/li-zhao}.

\bibitemdeclare{article}{lorido2014review}
\bibitem{lorido2014review}
\bibinfo{author}{Tania \surnamestart Lorido-Botran\surnameend},
  \bibinfo{author}{Jose \surnamestart Miguel-Alonso\surnameend} \&
  \bibinfo{author}{Jose~A \surnamestart Lozano\surnameend}
  (\bibinfo{year}{2014}): \emph{\bibinfo{title}{A review of auto-scaling
  techniques for elastic applications in cloud environments}}.
\newblock {\slshape \bibinfo{journal}{Journal of grid computing}}
  \bibinfo{volume}{12}(\bibinfo{number}{4}), pp. \bibinfo{pages}{559--592},
  \doi{10.1007/s10723-014-9314-7}.

\bibitemdeclare{misc}{k8-resource-optimizer/impl-public}
\bibitem{k8-resource-optimizer/impl-public}
\bibinfo{author}{Eddy~Truyen \surnamestart Matthijs~Kaminski\surnameend}
  (\bibinfo{year}{2024}):
  \emph{\bibinfo{title}{k8-scalar/k8-resource-optimizer}}.
\newblock
  \bibinfo{howpublished}{\url{https://github.com/k8-scalar/k8-resource-optimizer/}}.
\newblock \bibinfo{note}{[Accessed: 2024-01-25]}.

\bibitemdeclare{article}{moreschini2025ai}
\bibitem{moreschini2025ai}
\bibinfo{author}{Sergio \surnamestart Moreschini\surnameend},
  \bibinfo{author}{Shahrzad \surnamestart Pour\surnameend},
  \bibinfo{author}{Ivan \surnamestart Lanese\surnameend},
  \bibinfo{author}{Daniel \surnamestart Balouek-Thomert\surnameend},
  \bibinfo{author}{Justus \surnamestart Bogner\surnameend},
  \bibinfo{author}{Xiaozhou \surnamestart Li\surnameend},
  \bibinfo{author}{Fabiano \surnamestart Pecorelli\surnameend},
  \bibinfo{author}{Jacopo \surnamestart Soldani\surnameend},
  \bibinfo{author}{Eddy \surnamestart Truyen\surnameend} \&
  \bibinfo{author}{Davide \surnamestart Taibi\surnameend}
  (\bibinfo{year}{2025}): \emph{\bibinfo{title}{AI Techniques in the
  Microservices Life-Cycle: A Systematic Mapping Study}}.
\newblock {\slshape \bibinfo{journal}{Computing}}
  \bibinfo{volume}{107}(\bibinfo{number}{100}),
  \doi{10.1007/s00607-025-01432-z}.

\bibitemdeclare{article}{morris1991factorial}
\bibitem{morris1991factorial}
\bibinfo{author}{Max~D \surnamestart Morris\surnameend} (\bibinfo{year}{1991}):
  \emph{\bibinfo{title}{Factorial sampling plans for preliminary computational
  experiments}}.
\newblock {\slshape \bibinfo{journal}{Technometrics}}
  \bibinfo{volume}{33}(\bibinfo{number}{2}), pp. \bibinfo{pages}{161--174},
  \doi{10.1080/00401706.1991.10484804}.

\bibitemdeclare{inproceedings}{nikravesh2015towards}
\bibitem{nikravesh2015towards}
\bibinfo{author}{Ali~Yadavar \surnamestart Nikravesh\surnameend},
  \bibinfo{author}{Samuel~A \surnamestart Ajila\surnameend} \&
  \bibinfo{author}{Chung-Horng \surnamestart Lung\surnameend}
  (\bibinfo{year}{2015}): \emph{\bibinfo{title}{Towards an autonomic
  auto-scaling prediction system for cloud resource provisioning}}.
\newblock In: {\slshape \bibinfo{booktitle}{Proceedings of the 10th
  International Symposium on Software Engineering for Adaptive and
  Self-Managing Systems}}, \bibinfo{organization}{IEEE Press}, pp.
  \bibinfo{pages}{35--45}, \doi{10.1109/SEAMS.2015.22}.

\bibitemdeclare{misc}{bofmfn}
\bibitem{bofmfn}
\bibinfo{author}{Fernando \surnamestart Noguiera\surnameend}
  (\bibinfo{year}{2019}): \emph{\bibinfo{title}{Bayesian Optimization}}.
\newblock
  \bibinfo{howpublished}{\url{https://github.com/fmfn/BayesianOptimization}}.

\bibitemdeclare{article}{rodriguez2018container}
\bibitem{rodriguez2018container}
\bibinfo{author}{Maria~A \surnamestart Rodriguez\surnameend} \&
  \bibinfo{author}{Rajkumar \surnamestart Buyya\surnameend}
  (\bibinfo{year}{2018}): \emph{\bibinfo{title}{Container-based cluster
  orchestration systems: A taxonomy and future directions}}.
\newblock {\slshape \bibinfo{journal}{Software: Practice and Experience}},
  \doi{10.1002/spe.2660}.

\bibitemdeclare{inproceedings}{autopilot}
\bibitem{autopilot}
\bibinfo{author}{Krzysztof \surnamestart Rzadca\surnameend},
  \bibinfo{author}{Pawel \surnamestart Findeisen\surnameend},
  \bibinfo{author}{Jacek \surnamestart Swiderski\surnameend},
  \bibinfo{author}{Przemyslaw \surnamestart Zych\surnameend},
  \bibinfo{author}{Przemyslaw \surnamestart Broniek\surnameend},
  \bibinfo{author}{Jarek \surnamestart Kusmierek\surnameend},
  \bibinfo{author}{Pawel \surnamestart Nowak\surnameend},
  \bibinfo{author}{Beata \surnamestart Strack\surnameend},
  \bibinfo{author}{Piotr \surnamestart Witusowski\surnameend},
  \bibinfo{author}{Steven \surnamestart Hand\surnameend} \&
  \bibinfo{author}{John \surnamestart Wilkes\surnameend}
  (\bibinfo{year}{2020}): \emph{\bibinfo{title}{Autopilot: workload autoscaling
  at Google}}.
\newblock \bibinfo{series}{EuroSys '20}, \bibinfo{publisher}{Association for
  Computing Machinery}, \doi{10.1145/3342195.3387524}.

\bibitemdeclare{book}{saltelli2008global}
\bibitem{saltelli2008global}
\bibinfo{author}{Andrea \surnamestart Saltelli\surnameend},
  \bibinfo{author}{Marco \surnamestart Ratto\surnameend},
  \bibinfo{author}{Terry \surnamestart Andres\surnameend},
  \bibinfo{author}{Francesca \surnamestart Campolongo\surnameend},
  \bibinfo{author}{Jessica \surnamestart Cariboni\surnameend},
  \bibinfo{author}{Debora \surnamestart Gatelli\surnameend},
  \bibinfo{author}{Michaela \surnamestart Saisana\surnameend} \&
  \bibinfo{author}{Stefano \surnamestart Tarantola\surnameend}
  (\bibinfo{year}{2008}): \emph{\bibinfo{title}{Global sensitivity analysis:
  the primer}}.
\newblock \bibinfo{publisher}{John Wiley \& Sons}, \doi{10.1002/9780470725184}.

\bibitemdeclare{inproceedings}{schurman2009user}
\bibitem{schurman2009user}
\bibinfo{author}{Eric \surnamestart Schurman\surnameend} \&
  \bibinfo{author}{Jake \surnamestart Brutlag\surnameend}
  (\bibinfo{year}{2009}): \emph{\bibinfo{title}{The user and business impact of
  server delays, additional bytes, and HTTP chunking in web search}}.
\newblock In: {\slshape \bibinfo{booktitle}{Velocity Web Performance and
  Operations Conference}}.

\bibitemdeclare{article}{shahriari2016taking}
\bibitem{shahriari2016taking}
\bibinfo{author}{Bobak \surnamestart Shahriari\surnameend},
  \bibinfo{author}{Kevin \surnamestart Swersky\surnameend},
  \bibinfo{author}{Ziyu \surnamestart Wang\surnameend}, \bibinfo{author}{Ryan~P
  \surnamestart Adams\surnameend} \& \bibinfo{author}{Nando \surnamestart
  De~Freitas\surnameend} (\bibinfo{year}{2016}): \emph{\bibinfo{title}{Taking
  the human out of the loop: A review of bayesian optimization}}.
\newblock {\slshape \bibinfo{journal}{Proceedings of the IEEE}}
  \bibinfo{volume}{104}(\bibinfo{number}{1}), pp. \bibinfo{pages}{148--175},
  \doi{10.1109/JPROC.2015.2494218}.

\bibitemdeclare{inproceedings}{snoek2012practical}
\bibitem{snoek2012practical}
\bibinfo{author}{Jasper \surnamestart Snoek\surnameend}, \bibinfo{author}{Hugo
  \surnamestart Larochelle\surnameend} \& \bibinfo{author}{Ryan~P \surnamestart
  Adams\surnameend} (\bibinfo{year}{2012}): \emph{\bibinfo{title}{Practical
  bayesian optimization of machine learning algorithms}}.
\newblock In: {\slshape \bibinfo{booktitle}{Advances in neural information
  processing systems}}, pp. \bibinfo{pages}{2951--2959}.
\newblock
  \urlprefix\url{https://proceedings.neurips.cc/paper_files/paper/2012/file/05311655a15b75fab86956663e1819cd-Paper.pdf}.

\bibitemdeclare{inproceedings}{truyen-woc2020}
\bibitem{truyen-woc2020}
\bibinfo{author}{Eddy \surnamestart Truyen\surnameend}, \bibinfo{author}{Bert
  \surnamestart Lagaisse\surnameend}, \bibinfo{author}{Wouter \surnamestart
  Joosen\surnameend}, \bibinfo{author}{Arnout \surnamestart
  Hoebreckx\surnameend} \& \bibinfo{author}{C\'{e}dric~De \surnamestart
  Dycker\surnameend} (\bibinfo{year}{2021}): \emph{\bibinfo{title}{Flexible
  Migration in Blue-Green Deployments within a Fixed Cost}}.
\newblock In: {\slshape \bibinfo{booktitle}{Proceedings of the 2020 6th
  International Workshop on Container Technologies and Container Clouds}},
  \bibinfo{series}{WOC'20}, \bibinfo{publisher}{Association for Computing
  Machinery}, \bibinfo{address}{New York, NY, USA}, pp.
  \bibinfo{pages}{13--18}, \doi{10.1145/3429885.3429963}.

\bibitemdeclare{inproceedings}{venkataraman2016ernest}
\bibitem{venkataraman2016ernest}
\bibinfo{author}{Shivaram \surnamestart Venkataraman\surnameend},
  \bibinfo{author}{Zongheng \surnamestart Yang\surnameend},
  \bibinfo{author}{Michael~J \surnamestart Franklin\surnameend},
  \bibinfo{author}{Benjamin \surnamestart Recht\surnameend} \&
  \bibinfo{author}{Ion \surnamestart Stoica\surnameend} (\bibinfo{year}{2016}):
  \emph{\bibinfo{title}{Ernest: Efficient Performance Prediction for
  Large-Scale Advanced Analytics.}}
\newblock In: {\slshape \bibinfo{booktitle}{NSDI}}, pp.
  \bibinfo{pages}{363--378}.
\newblock
  \urlprefix\url{https://www.usenix.org/conference/nsdi16/technical-sessions/presentation/venkataraman}.

\bibitemdeclare{inproceedings}{verma2015large}
\bibitem{verma2015large}
\bibinfo{author}{Abhishek \surnamestart Verma\surnameend},
  \bibinfo{author}{Luis \surnamestart Pedrosa\surnameend},
  \bibinfo{author}{Madhukar \surnamestart Korupolu\surnameend},
  \bibinfo{author}{David \surnamestart Oppenheimer\surnameend},
  \bibinfo{author}{Eric \surnamestart Tune\surnameend} \& \bibinfo{author}{John
  \surnamestart Wilkes\surnameend} (\bibinfo{year}{2015}):
  \emph{\bibinfo{title}{Large-scale cluster management at Google with Borg}}.
\newblock In: {\slshape \bibinfo{booktitle}{Proceedings of the Tenth European
  Conference on Computer Systems}}, \bibinfo{organization}{ACM},
  p.~\bibinfo{pages}{18}, \doi{10.1145/2741948.2741964}.

\bibitemdeclare{inproceedings}{xi2004smart}
\bibitem{xi2004smart}
\bibinfo{author}{Bowei \surnamestart Xi\surnameend}, \bibinfo{author}{Zhen
  \surnamestart Liu\surnameend}, \bibinfo{author}{Mukund \surnamestart
  Raghavachari\surnameend}, \bibinfo{author}{Cathy~H \surnamestart
  Xia\surnameend} \& \bibinfo{author}{Li~\surnamestart Zhang\surnameend}
  (\bibinfo{year}{2004}): \emph{\bibinfo{title}{A smart hill-climbing algorithm
  for application server configuration}}.
\newblock In: {\slshape \bibinfo{booktitle}{Proceedings of the 13th
  international conference on World Wide Web}}, \bibinfo{organization}{ACM},
  pp. \bibinfo{pages}{287--296}, \doi{10.1145/988672.988711}.

\bibitemdeclare{article}{survey_buyya_2022}
\bibitem{survey_buyya_2022}
\bibinfo{author}{Zhiheng \surnamestart Zhong\surnameend},
  \bibinfo{author}{Minxian \surnamestart Xu\surnameend},
  \bibinfo{author}{Maria~Alejandra \surnamestart Rodriguez\surnameend},
  \bibinfo{author}{Chengzhong \surnamestart Xu\surnameend} \&
  \bibinfo{author}{Rajkumar \surnamestart Buyya\surnameend}
  (\bibinfo{year}{2022}): \emph{\bibinfo{title}{Machine Learning-based
  Orchestration of Containers: A Taxonomy and Future Directions}}
  \bibinfo{volume}{54}(\bibinfo{number}{10s}).
\newblock \doi{10.1145/3510415}.

\bibitemdeclare{inproceedings}{zhu2017bestconfig}
\bibitem{zhu2017bestconfig}
\bibinfo{author}{Yuqing \surnamestart Zhu\surnameend}, \bibinfo{author}{Jianxun
  \surnamestart Liu\surnameend}, \bibinfo{author}{Mengying \surnamestart
  Guo\surnameend}, \bibinfo{author}{Yungang \surnamestart Bao\surnameend},
  \bibinfo{author}{Wenlong \surnamestart Ma\surnameend},
  \bibinfo{author}{Zhuoyue \surnamestart Liu\surnameend},
  \bibinfo{author}{Kunpeng \surnamestart Song\surnameend} \&
  \bibinfo{author}{Yingchun \surnamestart Yang\surnameend}
  (\bibinfo{year}{2017}): \emph{\bibinfo{title}{Bestconfig: tapping the
  performance potential of systems via automatic configuration tuning}}.
\newblock In: {\slshape \bibinfo{booktitle}{Proceedings of the 2017 Symposium
  on Cloud Computing}}, \bibinfo{organization}{ACM}, pp.
  \bibinfo{pages}{338--350}, \doi{10.1145/3127479.3128605}.

\end{thebibliography}

\begin{appendices}

\section{\texttt{Values.yaml} file for TeaStore Deployment}
\label{sect:app-values}

\begin{lstlisting}[language=YAML, caption={Helm chart values for deploying TeaStore in K8s.}]
namespace: teastore
version: 1.2.0  #TeaStore version=1.2.0

persistenceReplicas: 1
persistenceCpu: 1000m
persistenceMemory: 1000Mi

webuiReplicas: 1
webuiCpu: 1000m
webuiMemory: 1000Mi

recommenderReplicas: 1
recommenderCpu: 1000m
recommenderMemory: 1000Mi

registryReplicas: 1
registryCpu: 500m
registryMemory: 1000Mi

dbReplicas: 1
dbCpu: 1000m
dbMemory: 1000Mi

authReplicas: 1
authCpu: 1000m
authMemory: 1000Mi

imageReplicas: 1
imageCpu: 1000m
imageMemory: 1000Mi
\end{lstlisting}

\section{Optimizer Configuration for TeaStore}
\label{sect:app-optimizer-configuration}
\begin{lstlisting}[language=YAML, caption={Configuration for k8-resource-optimizer targeting the TeaStore deployment.}]
nbOfIterations: 10
nbOfSamplesPerIteration: 6
charts:
  - name: teastore
    chartdir: charts/teastore-helm
slas:
  - name: silver
    chartName: teastore
    slos: 
      throughput: 0.5
      99th: 1000.0
    nbOfTenants: 10
    parameters:
      - name: persistenceCpu
        searchspace:
          min: 500
          max: 1125
          granularity: 125
        suffix: m
      - name: persistenceMemory
        searchspace:
          min: 512
          max: 1152
          granularity: 128
        suffix: Mi
      - name: webuiCpu
        searchspace:
          min: 500
          max: 1125
          granularity: 125
        suffix: m
      - name: webuiMemory
        searchspace:
          min: 512
          max: 1152
          granularity: 128
        suffix: Mi
      - name: recommenderCpu
        searchspace:
          min: 500
          max: 1125
          granularity: 125
        suffix: m
      - name: recommenderMemory
        searchspace:
          min: 512
          max: 1152
          granularity: 128
        suffix: Mi
      - name: dbCpu
        searchspace:
          min: 500
          max: 1125
          granularity: 125
        suffix: m
      - name: dbMemory
        searchspace:
          min: 512
          max: 1152
          granularity: 128
        suffix: Mi
		  #similar settings for other resource parameters
      
namespaceStrategy: NSPSLA
optimizer: bestconfig
utilFunc: teastore
outputDir: teastore-bestconfig
\end{lstlisting}

\section{Templating a String based parameter in an helm chart for K8-resource optimizer}
\label{app:string-param}
\begin{lstlisting}[language=yaml,caption={\texttt{values.yaml}}]
mySettingChoice: 1
\end{lstlisting}

\begin{lstlisting}[language=yaml,caption={\texttt{values.schema.yaml}}]
properties:
  mySettingChoice:
    type: integer
    enum:
      - 1
      - 2
    description: |
      1 = SimpleSetting
      2 = ComplexSetting
\end{lstlisting}

\begin{lstlisting}[language=yaml,caption={\texttt{deployment.yaml} file containing mySettingChoice parameter}]
# templates/deployment.yaml
apiVersion: apps/v1
kind: Deployment
metadata:
  name: myapp
spec:
  replicas: 1
  selector:
    matchLabels:
      app: myapp
  template:
    metadata:
      labels:
        app: myapp
    spec:
      containers:
        - name: myapp
          image: myapp:latest
          env:
            - name: MY_SETTING
              value: |
                {{ index (dict 1 SimpleSetting 2 ComplexSetting) .Values.mySettingChoice }}
\end{lstlisting}

\section{Implemented benchmark scenario}
\label{app:scenario}

\begin{table}[h!]
\centering
\begin{tabular}{|c|l|p{8cm}|}
\hline
\textbf{Step} & \textbf{Task} & \textbf{What happens} \\
\hline
0 & user & Picks a random user postfix number (e.g., user37, user58, etc.) \\
1 & index & Visits the homepage (GET /) \\
2 & loginPage & Visits the login page (GET /login) \\
3 & loginAction & Logs in with a username/password. Parses the response to find category links \\
4 & seeCategoryPage & Visits a random category page parsed from the login response \\
5 & seeProductPage & From the category page, picks a random product page \\
6 & addToCart & Adds the product to the cart (GET /cartAction?addToCart\&productid=X) \\
7 & seeCategoryPage2 & Visits another category page (repeat) \\
8 & seeProductPage2 & Visits another product page (repeat) \\
9 & addToCart2 & Adds another product to the cart \\
10 & getProfile & Visits the user's profile page (GET /profile) \\
11 & startAndLogout & Repeats homepage visit, then logs out (GET /loginAction?logout=) and clears cookies \\
\hline
\end{tabular}
\caption{Simulated User Flow: Step, Task, and Description. A waiting time of 0ms is set between tasks.}
\label{tab:step_task_what_happens}
\end{table}

\end{appendices}

\end{document}